# Innovations in Integrating Machine Learning and Agent-Based Modeling of Biomedical Systems

Nikita Sivakumar[1†], Cameron Mura[2†*], and Shayn M. Peirce[1]

[1] Department of Biomedical Engineering, University of Virginia, Charlottesville, VA, USA
[2] School of Data Science, University of Virginia, Charlottesville, VA, USA

last touched: 09 October 2022

*Correspondence: C. Mura; cmura@virginia.edu

[†]These authors contributed equally to this work.

Author contribution statement (for *Frontiers* submission system):
NS, CM, and SMP were each actively involved in the design and writing of this review article, from early drafting to final revision stages; the graphical illustrations/figures were created by NS, with input from CM and SMP.

Abbreviations, acronyms: ABM, agent-based model; AI, artificial intelligence; ANN, artificial neural network; BN, Bayesian network; CA, cellular automata; DNN, deep neural network; DR, diabetic retinopathy; EHR, electronic health record; FCM, fuzzy cognitive map or fuzzy *c*-means (depending upon context); GA, genetic algorithm; ML, machine learning; OOD, out-of-distribution; PDE, partial differential equation; RL, reinforcement learning; RWE, real-world evidence; SA, sensitivity analysis;

# CONTENTS

# 1. Abstract

Agent-based modeling (ABM) is a well-established computational paradigm for simulating complex systems in terms of the interactions between individual entities that comprise the system's population. Machine learning (ML) refers to computational approaches whereby algorithms use statistical methods to 'learn' from data on their own, i.e. without imposing any *a priori* model/theory onto a system or its behavior. Biological systems—ranging from molecules, to cells, to entire organisms, to whole populations and even ecosystems—consist of vast numbers of discrete entities, governed by complex webs of interactions that span various spatiotemporal scales and exhibit nonlinearity, stochasticity, and variable degrees of coupling between entities. For these reasons, the macroscopic properties and collective dynamics of biological systems are generally difficult to accurately model or predict via continuum modeling techniques and mean-field formalisms. ABM takes a 'bottom-up' approach that obviates common difficulties of other modeling approaches by enabling one to relatively easily create (or at least propose, for testing) a set of well-defined 'rules' to be applied to the individual entities (agents) in a system. Quantitatively evaluating a system and propagating its state over a series of discrete time-steps effectively simulates the system, allowing various observables to be computed and the system's properties to be analyzed. Because the rules that govern an ABM can be difficult to abstract and formulate from experimental data, at least in an unbiased way, there is a uniquely synergistic opportunity to employ ML to help infer optimal, system-specific ABM rules. Once such rule-sets are devised, running ABM calculations can generate a wealth of data, and ML can be applied in that context too—for example, to generate statistical measures that accurately and meaningfully describe the stochastic outputs of a system and its properties. As an example of synergy in the other direction (from ABM to ML), ABM simulations can generate plausible (realistic) datasets for training ML algorithms (e.g., for regularization, to mitigate overfitting). In these ways, one can envision a variety of synergistic ABM⇌ML loops. After introducing some basic ideas about ABMs and ML, and their limitations, this Review describes examples of how ABM and ML have been integrated in diverse contexts, spanning spatial scales that include multicellular and tissue-scale biology to human population-level epidemiology. In so doing, we have used published studies as a guide to identify ML approaches that are well-suited to particular types of ABM applications, based on the scale of the biological system and the properties of the available data.

## 2. INTRODUCTION

### 2.1 BIOMEDICAL SYSTEMS AND A DATA-DRIVEN MOTIVATION FOR MECHANISTIC MODELLING

Rapid advances in experimental methodologies now enable us to obtain vast quantities of data describing the individual entities in a population, such as single cells within complex, multicellular tissues, or individual patients in large-scale epidemiological systems. Hence, a growing focus of systems biology and biomedical research involves elucidating *patterns* in these large datasets—what are the associations between the discrete, individual entities themselves, and what are the mechanisms by which the behaviors, states and interactions of these autonomous agents contribute to broader-scale, population-wide outcomes? For example, single-cell RNA sequencing (Hwang et al., 2018; Potter, 2018; Kulkarni et al., 2019), single-cell proteomics (Irish et al., 2006; Marx, 2019), and flow cytometry (Wu et al., 2012; Wu et al., 2013; Chen et al., 2019; Argüello et al., 2020) provide snapshots of an individual cell's state at a single point in time. Yet, understanding the tissue–level and organ–level implications of these single-cell data requires a fundamentally different set of analytical approaches, with the capacity to spatiotemporally integrate potentially disparate data along at least two 'dimensions': across entire populations of entities, as well as across multiple scales (cellular $\rightarrow$ organismal). For instance, using single-cell RNA-Seq one can detect the presence and quantify the amount of RNAs in each of the cells in a tumor, but these data alone cannot illuminate *how* that particular collection of tumor cells, which undergo different behaviors (e.g., proliferation and migration) as dictated by their unique cellular states, contribute to tissue-level and organ-level outcomes such as angiogenesis and metastasis.

Epidemiological datasets have similarly expanded in recent years (Andreu Perez et al., 2015; Ehrenstein et al., 2017; Saracci, 2018). Many factors have driven this growth, including: (i) improved standardization and systematization of electronic health records (EHRs; (Andreu Perez et al., 2015; Booth et al., 2016; Casey et al., 2016; Ponjoan et al., 2019)), with concomitantly increased storage, more sophisticated cyberinfrastructure (e.g., cloud computing) and improved data-mining capacities, (ii) adoption of high-resolution, diagnostic medical imaging (Backer et al., 2005; Smith-Bindman et al., 2008; Andreu Perez et al., 2015; Preim et al., 2016), (iii) acquisition of genomic and other 'omics' big data, largely via next-generation sequencing (Thomas, 2006; Andreu Perez et al., 2015; Maturana et al., 2016; Chu et al., 2019), (iv) technologies such as wearable patient health sensors (Atallah et al., 2012; Andreu Perez et al., 2015; Guillodo et al., 2020; Perez-Pozuelo et al., 2021), and (v) acquisition of spatial and environmental data from geographical information systems (GIS; (Krieger, 2003; Rytkönen, 2004; Andreu Perez et al., 2015)). Despite these many technological advances, without a sound analytical and methodological framework to integrate and then explore these data it remains difficult to understand how, for example, certain lifestyle behaviors or healthcare policies might contribute to the spread of disease in a population of unique individuals. The challenges and unrealized potential of big data also hold true at the finer scale of physiological systems, from organs and tissues down to cellular communities, individual cells, and even at the subcellular scale. To handle biological big data, virtually all modern computational analysis pipelines employ machine learning (ML) methods, described below. While ML offers a powerful family of approaches for *handling* and *analyzing* big data, as well as drawing inferences, mechanistic insights and questions of causality are generally not as readily elucidated via ML; for this, we turn to mechanistic modeling.

### 2.2 OVERVIEW OF AGENT-BASED MODELING: GENERAL APPROACH, AND EXAMPLE APPLICATIONS

The quantitative determination and forecasting of *how* individual mechanisms contribute to system-level outcomes (known as "emergent properties") underpins much of basic research, and is also critical to applied areas such as creating targeted treatments, mitigating disease spread and, ultimately, guiding more informed healthcare policies. As described by Bonabeau (2002), these general types of problems—deciphering the global, collective behavior that emerges in a complex system, composed of a statistically large collection of (locally) interacting, individual components—are particularly amenable to the approach known as agent-based modeling (ABM). Thus, one can imagine synergistically integrating ABM and ML, leveraging the respective strengths of each: the mechanistic nature of ABMs is a potentially powerful instrument with which to analyze the (non-mechanistic, black-box) predictions that are accessible via ML. After briefly introducing ABM and ML, the remainder of this Review describes their potential synergy, including the relative strengths and limitations that have surfaced in recent studies that integrate these disparate approaches and apply them at various scales, from cellular to ecological.

ABM is now a well-established computational paradigm for simulating a system's population-level outcomes based on interactions between the individual entities comprising the population. Such an approach or ABM 'mindset' (Bonabeau, 2002) is necessarily bottom-up, and the general method has been applied to domains ranging from the dynamics of macroscopic systems, such as global financial markets (Bonabeau, 2002), to the microscopic dynamics of mRNA export from the cellular nucleus (Soheilypour and Mofrad, 2018). ABMs simulate spatially-discrete, autonomous individuals, or 'agents', that follow relatively simple 'rules' across a series of discrete time-steps. Rules are formulated to describe the discrete, well-defined, individual behaviors that a single agent can enact at a given time-step depending upon (and possibly in response to) both its own state and its local environment; also, rules can be either probabilistic or deterministic, and can flexibly take into account prior agent states, simulation time, and other system-specific parameters. The agents—which can represent proteins, biological cells, individual organisms, or really any definable entity (e.g., individual traders in a stock market)—exhibit specific behaviors over time, and the collection of these behaviors (e.g., patterns of pairwise interactions) gives rise to population-level outcomes in the system/ensemble, such as in embryonic development (Longo et al., 2004; Robertson et al., 2007; Thorne et al., 2007a), blood vessel growth (Peirce et al., 2004; Thorne et al., 2011; Walpole et al., 2015), disease progression within a given organism (Martin et al., 2015; Virgilio et al., 2018), or infectious disease spread among subsets of organisms in a population (Cuevas, 2020; Rockett et al., 2020). As noted in a recent study that developed a "biological lattice-gas cellular automaton" (BIO-LGCA) for the collective behavior of cellular entities in complex multicellular systems (Deutsch et al., 2021), classical, on-lattice ABMs and cellular automata (CA) are similar approaches. In particular, CAs also simulate phenomena via a grid-based spatial representation, but often do not explicitly consider interactions between agents (in that way, CAs may be viewed as simplified forms of the ABM). For a recent introductory overview of ABMs, from a mathematical perspective and with an illustrative application to a three-state disease transmission model, see Shoukat & Moghadas (2020).

The discretized nature of ABMs, in time and space, allows them to capture the temporal stochasticity and spatial heterogeneities that are inherent to most complex dynamical systems, biological or otherwise. The variabilities that arise from stochasticity and heterogeneity[1] can be handled in ABM simulations in a manner that is numerically robust (e.g., to singularities, divergences) and, thus, capable of emulating how real biological processes may progress towards potentially different outcomes (i.e., non-deterministic behavior), based on (i) heterogeneity among the unique agents in a population, (ii) stochasticity at the level of individual agents (i.e. inherent variability stemming from differential responses of each individual agent), or (iii) variability in the environment and the coupling of agents to that potentially dynamic environment (i.e., spatial inhomogeneities). Moreover, ABMs can quantitatively predict numerous outcomes for a dynamical system that may be difficult or impossible to quantify experimentally, at least with sufficient spatial and temporal resolutions.[2] ABMs can also be leveraged to predict system behaviors in response to a wide range of different perturbations and initial conditions, allowing for a more comprehensive understanding of biomedical systems than is accessible via experimentation alone. Another key benefit of ABM is that its accuracy as a modeling approach does not suffer from the requirement of average-based assumptions for a system (fully-mixed, mean-field approximations, etc.), in contrast to partial differential equation (PDE)—based approaches or other modeling frameworks that treat a system as a smooth continuum, free of singularities and irregularities. Combined with methodological approaches such as sensitivity analysis (ten Broeke et al., 2014; ten Broeke et al., 2016; Ligmann-Zielinska et al., 2020), these attributes make ABMs particularly useful tools for examining the dependencies of population-level phenomena on the behaviors and interactions of the individuals that comprise the population (e.g., using ABM simulations to map a 'response surface', in terms of some underlying set of features/explanatory variables (Willett et al., 2015)).

---

[1] The term 'stochastic' generally refers to temporal events (e.g., "a stochastic *process*"), while 'heterogeneity' occurs as a spatial or population-wide phenomena (e.g., "a heterogeneous *population*"). Usage of the terms in some fields can be somewhat linked, though the concepts are distinct (Gregg et al., 2021): (i) *heterogeneity* can be regarded as variance in a specific feature across a population or spatial domain, while (ii) *stochasticity* means differential responses from an individual agent under identical stimuli (i.e., it applies in the time domain).

ABMs have been used extensively to model multicellular processes, such as tissue patterning and morphogenesis (Robertson et al., 2007; Thorne et al., 2007a; Thorne et al., 2007b; Taylor et al., 2017), tumorigenesis (Wang et al., 2007; Gerlee and Anderson, 2008; Zangooei and Habibi, 2017; Oduola and Li, 2018; Warner, 2019), vascularization (Walpole et al., 2015; Walpole et al., 2017; Bora et al., 2019), immune responses (An, 2006; Bailey et al., 2007; Woelke et al., 2011; Xu et al., 2018), and pharmacodynamics (Hunt et al., 2013; Cosgrove et al., 2015; Kim et al., 2015; Ji et al., 2016; Gallaher et al., 2018). In the field of epidemiology, ABMs have also been used to represent human individuals in a population in order to study infectious disease transmission and to create simulations of disease spread in a cohort of individuals over time (Marshall and Galea, 2015; Çağlayan et al., 2018; Tracy et al., 2018; Cuevas, 2020; Rockett et al., 2020); note that in epidemiological, geographic, social, economic and some other settings, the terms ABM and 'microsimulations' are often used interchangeably, though they are distinct approaches (Heard et al., 2015; Bae et al., 2016; Ballas et al., 2016). Like CAs, microsimulations also simulate individual entities in a discretized space over discrete timesteps, and do not have individual agents interact.[3] While ABMs have been applied widely and fruitfully in biomedical research, they are not without recognized limitations. For example, the rules that govern ABMs can be difficult to abstract and formulate from experimental data alone, at least in a minimally-biased way. In addition, running ABMs can be computationally expensive, and selecting statistical measures that accurately and meaningfully summarize stochastic outputs can be challenging (ten Broeke et al., 2016). Some of these constraints and limitations may be alleviated by leveraging machine learning (ML) approaches as part of an ABM pipeline.

## 2.3  What is ML, and How Is It Useful in Studying Biomedical Systems?

Machine learning is a vast set of approaches whereby algorithms use statistical formalisms and methods to 'learn' from data on their own—that is, without being explicitly programmed to do so. In ML, whatever relationships, patterns or other associations that may latently exist in a body of data are gleaned from the data, without requiring an *a priori* theory or model to specify the details of the possible relationships in advance. Given that, note that some general *form* for a model must be posited—e.g., we assume datasets can be fit by a linear function, can be represented by a neural network, exist as natural clusters of associations, or so on; how these unavoidable assumptions manifest is known as the *inductive bias* of an ML model or learning algorithm. The next subsection (2.3.1) introduces some foundational ML concepts and terminology via a basic description of *supervised learning*; then, the remaining subsections consider the issue of evaluating ML models (2.3.2) and, finally, a broader perspective of ML more generally (2.3.3), including a taxonomic overview of ML approaches (supervised, unsupervised, etc.).

---

[2] However, note that predicting 'unobservable' data is a fraught endeavor: it is far from straightforward to assess the validity of models or simulations without "ground truth" data from real, empirically well-characterized systems with which to compare.

[3] CAs, ABMs and microsimulations share many similarities, though they are not equivalent. All of these approaches are bottom-up and discretized (spatiotemporally, and at the level of individual agents that enact their own decision-making). However, whereas the entities in a microsimulation are rather autonomous and updated stochastically (essentially corresponding to a population of non-interacting agents), agent⋯agent social interactions are pivotal in ABM; indeed, it is this social coupling that underpins the emergence of collective, population-wide behaviors in ABMs.

### 2.3.1 FOUNDATIONAL ML CONCEPTS, THROUGH THE LENS OF SUPERVISED LEARNING

Most broadly, every ML project begins with a question, data, and a model. Armed with a model, and as cogently put by Alpaydin (2021), "*it is not the programmers anymore but the data itself that defines what to do next*". How this learning is achieved can be understood by considering ML, most generally, as a way to determine a function that optimally maps a dataset (captured as a set of independent variables, often termed 'features') to a set of results/outcomes (dependent variables). That is, we aim to find a function $\mathcal{F}$ that maps the data to the 'results', $\{d\} \mapsto \{r\}$, where the 'map' can be as simple as a linear model (i.e., weighted sum of linear terms) or as complex as a deep neural network (DNN) in the context of ML via neural networks (NNs). In the latter case of artificial NNs (ANNs), such as in deep learning, the many thousands of 'hyperparameters' (synaptic weights between neurons, activation thresholds, biases, etc.) that define/parameterize a DNN model are learned via backpropagation algorithms, and it is the DNN's pattern of connectivities and weights that defines the functional map.[4] We generally seek a map that is 'optimal' in terms of minimizing the error between predicted outcomes and preexisting outcomes that are already known (as a 'ground truth') for a given subset of data (the 'training set'). The 'learning' part of machine learning is essentially the iterative approximation of parameters to optimize a model against an objective function (again, in many ML methods a functional form/class is assumed in advance, when deciding to model data via one approach or another). This training process, in turn, simply means finding a set of weights/parameters that optimizes the map's fitness, which corresponds to minimizing the model's prediction error; fundamentally, these 'weights' can be as simple as the numerical coefficients of the variable terms in the optimization function (complexity stems from there being myriad such parameters), and, as an example objective function commonly used in linear regression, minimizing the mean-squared error is equivalent to maximizing the likelihood of the (normally-distributed) data being observed under our (optimized) linear model. Reaching this point of having optimized the parameters, we are said to have a "*trained model*"; the second stage of an ML project is generally the "*inference stage*", wherein a trained model is then utilized to make predictions. This latter stage involves applying the trained model across new/unseen data (i.e., individual data items), which can take the form of either test datasets (while still developing/validating the model) or else real-world data (if the model is being deployed for production purposes, e.g. as part of a computational pipeline for automated tagging of chest X-rays in a clinic).

---

[4] That such a functional map (again, by 'map' we mean an arbitrary, nonlinear input→output mapping) is realizable is assured for generic feedforward NN architectures by the *universal approximation theorem*, which rigorously proves that there exists some NN with the capacity to model/approximate an arbitrary, continuous function to any desired level of accuracy, given enough neurons (these *hidden units* can be stacked to some *depth* [number layers] or arrayed across some *width* [number neurons per layer]). Haykin's text (2009) provides an authoritative treatment of this and related topics, including the interplay between function approximation and (i) the robustness of a NN's generalization properties (e.g., how overfitting/overtraining limits a NN's ability to accurately estimate a broader neighborhood of the functional space [i.e., *generalize*], as the NN effectively memorizes [rather than generalizes] an input/output map, yielding a 'brittleness' of sorts), (ii) how much input data suffices for 'good' generalization to be possible, and (iii) the *curse of dimensionality*, which, in this context of training NN models, describes how dataset limitations pose exponentially greater difficulties for model training as the dimensionality of a problem grows (an acute issue in biomedical systems, where data may be sparse in terms of both volume and in how representatively they sample the system of interest).

### 2.3.2 ASSESSING ML MODELS: AN IMPORTANT AND TRICKY TOPIC

Model training, selection, and evaluation are critical stages of any ML development pipeline, including when considering potential integration with an ABM framework. Even in broad, highly generic terms—considering, e.g., classification problems versus regression problems, ANN-based ML versus other types of ML, and so on—interrelated topics such as *model selection* and *performance evaluation* of ML models comprise an entire field unto itself. For a thorough treatment of this topic, we suggest Raschka's recent work (2018). Here, we simply note a few points. First, model training (i.e., the core *learning* part of the ML pipeline) and model selection are generally not inseparable issues—e.g., when constructing 'splits' of an overall dataset into *training* (say 80% of the data) and *test* (say 20% of the data) sets, and then further carving out a *validation* subset (disjoint from the rest of the training set). How an ML approach is evaluated greatly depends on the form and complexity of its model, and an evaluation approach that may be suitable for regression-based or other 'shallow' learning methods, both in terms of the ratios of splits as well as *how* data are assigned to training/validation/test sets, may not be as justifiable for more complex models, such as DNNs. Second, model development and performance is closely linked to the key goal of mitigating overfitting/underfitting; optimally achieving that balance yields the most successful model, with the lowest "*generalization error*" (i.e., accuracy of future predictions with unseen data, particularly data-points that lie distant from the training set). Third, (i) the general *approaches* (e.g., cross-validation, Bayesian model selection, etc.) and (ii) the *types of evaluation metrics* (e.g., a log-loss function for classification tasks, a mean-squared error for regression, etc.) can vary greatly with the *type of ML* being performed (e.g., ANN versus non-ANN). In short, when considering ML approaches for integration with an ABM framework, care must be taken so as to not inadvertently confound the various approaches used in training, testing, validating, and otherwise evaluating the training and performance of ML models, including such issues as whether the models are ANN– versus non-ANN–based. As with all applications of ML, caution must be exercised in taking into account the type of ML when formulating validation and evaluation strategies.

### 2.3.3 A BROADER, TAXONOMIC PERSPECTIVE ON ML

While that which precisely distinguishes an algorithm as a *machine learning* algorithm (cf. statistical modeling, for example) is debatable, here we consider ML algorithms as broadly defined by Mitchell's criterion (Mitchell, 1997): an algorithm is said to 'learn' if it improves its performance, $\mathcal{P}$, with respect to a task, $\mathcal{T}$, via execution of some computational processes, $\mathcal{E}$. ML algorithms broadly fall into four main types, depending chiefly on the role of labeled data in the training and learning process (note that the distinctions between some of these types are blurred): *Supervised learning*, *Unsupervised learning*, *Semi-supervised learning*, and *Reinforcement learning*. (i) *Supervised learning* algorithms infer relationships between independent and dependent variables by applying a model that was *trained* against prior data of known type (Bhavsar and Ganatra, 2012; Singh et al., 2016); these 'ground-truth' data consist of (accurately) 'labeled' samples which can be split in various ways (e.g., into 'training', 'test' and 'validation'/'development' sets) as part of the model-training regime. (ii) In *unsupervised learning*, an ML algorithm autonomously identifies patterns, trends or 'groupings' (clusters) in a dataset, with zero human intervention in the form of prior knowledge about the 'correct' associations—i.e. only unlabeled data (unannotated by human experts) are available for use (Gentleman and Carey, 2008; Kassambara, 2017). (iii) *Semi-supervised learning*, which can be employed when large volumes of data are available but only a small fraction of it is (correctly) labeled, is characterized by training sets that contain both labeled and unlabeled data, with a preponderance of the latter. Here, we consider "informed ML" (von Rueden et al., 2021), or "expert knowledge–driven" ML, as a form of semi-supervised learning, wherein the modeler may manually adjust model weights based on known behaviors and properties of the system. (iv) In *reinforcement learning*, the ML process acts via a set of rules ('policies') and series of system 'states' (akin to ABM), and is rewarded or punished based on the result of a 'move'; the ML algorithm evolves the system via state transitions so as to maximize rewards in a given environment (Sutton, 1992; Kaelbling et al., 1996; Kulkarni, 2012), and in that way it 'learns'. A rich variety of learning algorithms fall within these broader categories, as summarized by the sample inventory of ML approaches in Table 1. While that Table is not comprehensive, each of the ML methods listed there has been widely applied to study and analyze various types of biological systems (proteins, networks, tissues, etc.).

Numerous useful reviews of ML in the biosciences have appeared in recent years, including the revolution in learning via DNNs (i.e., deep learning [DL]). As but one example, Ching et al. (2018) have thoroughly reviewed the challenges and opportunities posed by applications of ML to big data in the biosciences. While focused on deep learning (LeCun et al., 2015; Goodfellow et al., 2016; Tang et al., 2019), many of the principles of that review apply to ML more broadly, e.g., viewing deep networks as analogous to ML's classic regression methods, but sufficiently generalized so as to allow for nonlinear relationships among features.[5] In the biomedical realm, ML has emerged as a powerful and generalized paradigm for integrating data to classify and predict phenotypes, genetic similarities, and disease states within various biological processes; here, again, numerous reviews are available (Chicco, 2017; Park et al., 2018; Serra et al., 2018; Jones, 2019; Nicholson and Greene, 2020; Su et al., 2020; Peng et al., 2021; Tchito Tchapga et al., 2021).

### 2.4 LIMITATIONS OF ML, AND POTENTIAL SYNERGIES OF INTEGRATING ABM/ML FOR BIOMEDICAL SYSTEMS?

Notwithstanding its many major successes, there are a few significant limitations associated with the application of ML to biology. Creating an accurate and robust ML model requires large amounts of experimental data, such as patient data, cellular-scale measurements, 'omics' data, etc.; such data may be challenging to obtain for various reasons, including measurement inaccuracies, inherent sparsity of datasets, or other concerns such as paucity of health data stemming from privacy policies. Perhaps most fundamentally vexing for modelers, ML architectures and algorithms generally pursue optimality criteria in a manner that yields "black-box" solutions, and unfortunately the internal mechanisms that may link predictor and outcome variables remain unknown; thus, ML algorithms often do not illuminate the causal mechanisms that underlie system-wide behavior (in the language above, we have found a map $\mathcal{F}$, but with no explanatory basis for it). For these types of reasons, "explainable A.I." has become a highly active research area (Jiménez-Luna et al., 2020; Confalonieri et al., 2021; Vilone and Longo, 2021). An interesting notion arises if we consider (i) the black-box property of ML methods, in tandem with (ii) the bottom-up, mechanistic design of ABMs (e.g., in terms of their discrete, well-defined rule-sets), and view it through the lens of (iii) the theory of causal inference (Pearl, 2000). Namely, ABM rule-sets are essentially collections of low-level *causal mechanisms* (i.e., *structural causal models*). Therefore, might it be possible to synergistically apply the data from ABM simulations to examine (high-level) ML-based predictions (i.e., hypotheses)? And, could doing so help 'unwind' the "ladder of causation" (Bareinboim et al., 2022), or *causal hierarchy*, that is induced by the ABM's set of causal mechanisms and that, at least loosely speaking, might underpin the "inner workings" (black-box) of the corresponding ML model? These types of questions can be explored by judiciously integrating ML and ABM.

---

[5] Note that DL's many successes in recent years have made it such a juggernaut that the term is often treated as being synonymous with ML; however, that is certainly not the case, and DNNs are but one form of artificial NNs (see below), which in turn belong to the vast family of ML methods.

With their respective strengths and limitations, the ABM and ML methodologies can be viewed as complementary approaches for modeling biological systems—particularly for systems and problems wherein the strengths of one type of methodology (ABM or ML) can be leveraged to address specific shortcomings of the other. For example, ML algorithms (e.g. DNNs) are often criticized because their predictions are arrived at in a black-box manner; in addition, most supervised algorithms require large amounts of accurately labeled training data, and overfitting is a common pitfall in many ML approaches. ABMs, on the other hand, (i) are built upon explicit representations/formulations of the precise interactions between system components (these rules are 'low-level', and thus relatively easily formulated), and (ii) can easily generate, via suites of simulations, large quantities of output data. Similarly, while the creation of rules in ABMs is frequently accomplished by manual and subjective curation of the literature, which can lead to a biased or oversimplified abstraction of the true biological complexity, ML approaches such as reinforcement learning can be used to computationally *infer* optimal rule-sets for agents and their interactions. Thus, there is a naturally synergistic relationship between these pairs of relative strengths and weaknesses of ABM and ML.

Inspired by this potential synergy, the remainder of this Review highlights some published studies that have integrated ML with ABMs in the following ways, in order to create more advanced and accurate computational models of biological systems, at both the multicellular and epidemiological scales:

- **Learning ABM Rules via ML**: Reinforcement learning and supervised learning methods can be used to infer and refine agent rules, which are critical inasmuch as these rules are applied at each discrete time step and, thus, largely define the ABM.
- **Parameter Calibration and Surrogate Models of ABMs**: Stochastic optimization approaches, such as genetic algorithms and particle-swarm methods, can be used to calibrate ABM parameters. Supervised learning algorithms can be trained to create surrogate models of an ABM, which also aids in calibration and in mitigating the computational costs of having to execute numerous ABMs.
- **Exploring ABMs via ML**: ML methods can help explore the complex, high-dimensional parameter space of an ABM, in terms of sensitivity analysis, model robustness, and so on.

As an overview of the high-level organization of this Review, Figure 1 schematizes how individual studies, reported in the recent literature, have integrated ML into each step of formulating and analyzing an ABM. We emphasize that our present Review is by no means comprehensive: we have merely focused on specific examples of the above types of integrations. Several prior reviews have described how ML can be leveraged in computational modeling, e.g. by Alber et al. (2019) and by Peng et al. (2021). In addition, the idea of synergistically integrating ML and ABM (Figure 2) has existed since at least Rand's (2006) early report, and includes more recent works such as by Giabbanelli (2019), Brearcliffe & Crooks (2021), and Zhang et al. (2021). The remainder of our present Review focuses more on the utility of ML within ABMs, and attempts to offer some guiding principles on how and when these integrations are feasible in simulating different scales of biology.

## 3. Using ML to Derive and Determine Agent Rules

### 3.1 Overview and Motivation: Where Do ABM 'Rules' Come From?

An ABM's rules define the autonomous actions that an agent can perform as a function of its state and in response to changes in its local environment. For instance, a cell may undergo apoptosis if it experiences sustained hypoxia, or a healthy individual may be infected with a virus when in close proximity to an infected individual. Note that the words 'can' and 'may' occur in the previous sentences because an ABM's rules are defined probabilistically. Rules link cause-and-effect in a manner that is enacted by individual agents/entities (molecules, cells, human individuals, etc.) in the population under consideration. Traditionally, these rules are manually generated by the modeler, who must curate and interpret empirical data describing the system, and synthesize that with expert opinions and/or dogma in the literature. In an ABM, the rule-set is only validated after ABM simulations have been run and predictions are compared to independent experimental data, or to a validation dataset. Hence, a common criticism of the traditional ABM rule-generation process is that there is inherent subjectivity on the part of the model-builder that could introduce bias in the rules, thereby skewing the biological relevance of its downstream results and predictions.

To overcome this potential issue, recent ABMs have begun leveraging ML to computationally determine—in a less ad hoc and heuristic manner—the rules governing agent behaviors based on an agent's spatial environment at a given time-step. Instead of manually-generated rules, which could be unwittingly biased towards a particular set of predictions that are not statistically representative of the target population or system behavior at large, ML algorithms can learn the rules, parameterizations, and so forth more objectively—by examining experimental data or by applying fundamental mathematical relationships (Figure 3); indeed, this "learn from the data" capacity stems directly from the roots of ML in information theory and statistical learning (Hastie et al., 2009).

### 3.2 Supervised Learning to Develop Agent Behaviors

Supervised learning algorithms have been leveraged in epidemiological ABMs to define agent behaviors in simulations of the spread of both infectious and non-communicable diseases. Indeed, supervised learning can be useful in ABM rule generation because of the capacity to 'learn' agent rules from labeled datasets that map agent features to agent behaviors under systematically varying conditions or circumstances. For example, a microsimulation (Day et al., 2013) of diabetic retinopathy (DR) in a cohort of individuals used a multivariate logistic regression algorithm to help build the rules that determine when each human agent will advance to the next stage of DR, based on features such as age, gender, duration of diabetes, current tobacco use, and hypertension. Instead of manually estimating DR stage advancement probabilities from the literature, these rules were computationally learned by training a multivariate logistic regression model on a dataset describing 535 DR patients. The logistic regression algorithm learned a function relating individual patient features to the probability of DR stage advancement, and at the beginning of every simulated year in the ABM this function was used to determine whether each human agent would advance to the next stage of DR. This approach showed that a simulated cohort of 501 patients had no significant differences from an actual live-patient cohort. Moreover, the logistic regression method was useful in identifying key predictors of DR stage advancement (Day et al., 2013). Finally, note that this example illustrates the general principle that regression models are highly applicable to constructing rules when large volumes of patient data are available.

Another study (Alexander Jr et al., 2019) evaluated multiple supervised learning methods to predict, in the context of an ABM platform, individual DR patient responses to pregabalin, a medication that targets the gabapentin receptor and which is used to treat several conditions, including diabetic neuropathy. The study found that 'ensemble' methods that combined several 'instance-based' learning methods, including supervised *k*-nearest neighbors and fuzzy *c*-means, yielded the highest classification accuracy (Alexander Jr et al., 2019).

Much recent effort in biomedical informatics has focused on developing approaches to automatically and systematically extract and infer statistically rigorous new information—so-called "real-world evidence" (RWE)—from primary data sources such as electronic health records (EHRs). The general aims of such efforts are manifold, including discovery of new uses for drugs already known to be safe and efficacious (an approach known as 'repurposing') and, ultimately, to reach high-confidence, clinically-actionable recommendations (e.g., a particular drug for a specific indication), ideally in a personalized, 'precision medicine' manner. A potentially synergistic interface can be found between ABMs and RWE-related studies, for example by using raw (low-level), patient-derived data to both develop ABMs (define rule-sets, parameterizations, etc.) and also deploy them for predictive purposes. For instance, real-world data about the spread of COVID-19 in hospitals and other settings have been used to develop and deploy ABMs for use in optimizing policy measures and exploration of other epidemiological questions (Gaudou et al., 2020; Hinch et al., 2021; Park and Sylla, 2021); also broadly notable, recent ABM-based studies of "information diffusion" have been used in the development of advanced community health resources (Lindau et al., 2021) and to examine how "medical innovation" might propagate among specific communities, such as cardiologists (Borracci and Giorgi, 2018).

### 3.3 EXPERT KNOWLEDGE-DRIVEN SUPERVISED LEARNING APPROACHES FOR ABMS

Other studies have demonstrated the utility of ML algorithms in ABM rule generation, even when there is limited available training data. Some studies train supervised learning algorithms on available data and augment the learned functions with expert knowledge. Bayesian networks (BNs), for example, are a common supervised learning algorithm that is paired with expert knowledge. First, the BN is trained on datasets to determine conditional probabilities of a certain event occurring based on predictor values, such as the probability of a sick individual infecting a healthy individual given the physical distance between them. Then, in a type of approach that has been termed "informed ML" (von Rueden et al., 2021), domain experts can subjectively adjust these learned probabilities based on experience and published literature. One study (Abdulkareem et al., 2019) used this approach to determine human agent rules in a previously developed ABM of cholera spread in Kumasi, Ghana. That work (Figure 3) compared four different BNs, trained with varying combinations of survey data and expert opinion support, to define a rule on whether a human agent would decide to use river water based on varying levels of (i) visual pollution, (ii) media influence, (iii) communication with neighbors, and (iv) past experience. The ABM was found to be most accurate when the BNs combined low-level data with expert knowledge; this is somewhat unsurprising, as the available training data were from a limited number of participants that did not holistically represent the modeled population. Moreover, the study found that a "sequential learning" approach further improved the accuracy of the ABM. Sequential learning refers to training the BN in an 'online' manner, simultaneously with the ABM simulation, such that the BN is re-trained on data that are generated during the ABM simulation. This study shows not only that Bayesian networks are a viable learning algorithm to incorporate expert opinion into ABMs when the available training data is limited, but also that the feedback process in sequential learning can further improve the accuracy of a learning algorithm by utilizing data generated by the ABM simulation. In another study, (Augustijn et al., 2020) took an alternative approach to determining human agent water use in the aforementioned cholera–spread ABM: that work trained decision trees on the same training dataset as the earlier study (Abdulkareem et al., 2019) in order to determine whether an agent will use river water based on the same predictor variables. The decision tree scheme differs from the BN approach in that the decision tree does not require expert opinion or sequential learning, and instead derives (novel) agent rules from scratch by determining a tree-like model/path of how each agent considers the predictor variables to arrive at a decision regarding usage of river water. The decision tree-based approach yielded ABM predictions with different numbers of infected individuals (Augustijn et al., 2020). This discrepancy could be anticipated because, as outlined in Figure 4, the two different ML integrations led to two fundamentally different rulesets, thus affecting the emergent properties/outcomes of the system. That these different integrations of ML in these two studies yielded different ABM results underscores the importance of testing multiple ML approaches and integration strategies in order to assess which method will yield the highest accuracy ABMs for the particular systems being examined.

Supervised learning approaches that are expert knowledge–driven have also been applied in studies that integrate artificial feed-forward neural networks (ANNs) into ABMs of multicellular systems. As the predecessors of today's deep NNs, information processing in ANNs is inspired by the hierarchical, multi-layered, densely-interconnected patterns of signaling and information flow between layers of neurons in the human brain. In an ANN, each neuron (or 'hidden unit') processes input variables, e.g. via a linear summation, and 'decides' how to pass this information on to the next neuron (downstream), the decision being based on whether or not the computed numerical values exceed an 'activation threshold'. (The foundations of NNs are treated in the classic text by Haykin (2009).) Ideally, an ANN's input variables capture salient features about a system in terms of its dynamics, local environment, and so on; non-numerical information (e.g., categorical data) can be captured as input via a process known as feature encoding. Also, note that the activation function and the neuron's input-combining functionality can range from relatively simple (e.g., a binary step function or a weighted linear combination of arguments) to more sophisticated forms, such as (i) those based on the hyperbolic tangent (or the similar logistic function, both of which sigmoidally saturate), or (ii) the more recent piecewise-linear 'rectified linear units' (ReLU), which are found to generally work well in training DNNs (Glorot et al., 2011).

As an example, one study incorporated an expert knowledge-based feed-forward ANN to determine cellular behavior based on environmental conditions in an ABM of tumor growth (Gerlee and Anderson, 2007). Within the ABM, each cellular agent encoded an ANN that decided cell phenotype based on inputs describing a cell's local environment, such as the number of cellular neighbors, local oxygen concentration, glucose concentration, and pH (Gerlee and Anderson, 2007). Each ANN processed these inputs to select one from a limited number of discrete phenotypic responses, such as proliferation, quiescence, movement, or apoptosis. Also, in that work the connection weights and activation thresholds of each neuron were manually set, thus 'tuning' the ANN such that overall cellular behavior resembled that of cancer cells (i.e., a certain percentage of cells in the population had each of the output phenotypes), instead of training the ANN on actual cellular data. As cells proliferated, they implicitly passed on their ANN to successive generations. Genetic mutations were incorporated into the simulation model by introducing random fluctuations in the ANN weights and thresholds when passed on to daughter cells (Gerlee and Anderson, 2007). These simulated genetic mutations allowed the authors to study clonal evolution in tumors and the environmental factors that contribute to the emergence of the glycolytic phenotype—a cellular state characterized by upregulated glycolysis, and which is known to increase the invasiveness of a tumor (Gerlee and Anderson, 2007; 2008).

The above ANN framework was incorporated into follow-up studies aimed at modeling drug delivery and hypoxia. Those further studies increased the complexity of the cellular ANN by adding growth and inhibitory factors as inputs (Kazmi et al., 2012a; Kazmi et al., 2012b), and also by introducing infusion of a bioreductive drug into the ABM; these studies explored the effects of protein binding on drug transport (Kazmi et al., 2012a; Kazmi et al., 2012b). Other studies used a similar ANN architecture to pinpoint effects of hypoxia on tumor growth (Al-Mamun et al., 2014), and explored the efficacy of a chemotherapeutic agent, maspin, on tumor metastasis (Al-Mamun et al., 2013; Al-Mamun et al., 2016). Overall, this design scheme—i.e., 'embedding' ANNs into the agent entities of an ABM—illustrates an intriguing and creative type of synergy that is possible when integrating ML and ABM–based approaches.

### 3.4 Reinforcement Learning: Determining Agent Behavior in Multicellular ABMs

Another type of learning algorithm used in multicellular ABMs is reinforcement learning (RL), which learns cellular behaviors as 'policies' that maximize a (cumulative) reward based on the surrounding environment and transitions of the system from one 'state' to the next (Table 1). Conceptual similarities between RL and ABM rule-sets run deep: the RL approach can be largely viewed as being a type of agent-based Markov decision process (Puterman, 1990). The key elements of this approach are four interrelated concepts: (i) the *state* that is occupied by an agent at a given instant (e.g., a cell can be in a 'quiescent' versus 'proliferating' state); (ii) the *action* which an agent can take (e.g., apoptose versus divide); (iii) a probabilistic *policy map*, specifying the chance (and rewards) of transitions between a given combination of states and actions (call it $\{s_i, a\}$) to a new state, $s_{i+1}$ (in other words, the conditional probability of taking action $a$ and thus adopting state $s_{i+1}$, while in state $s_i$); and (iv) the notion of a *reward*, *value* or *return* (these interrelated quantities can be treated as equivalent for present purposes), which is computed both instantaneously, for incremental state transitions $i \rightarrow i + 1$, as well as cumulatively (a global reward value, for the entire/completed process; ultimately, RL methods seek to maximize this latter quantity). The reward function can be formulated by the modeler to promote known/expected cellular behaviors, such as elevated metabolism in the presence of glucose or contact inhibition when surrounded by cells (Kaelbling et al., 1996; Kulkarni, 2012).

Recent studies of multicellular systems (Zangooei and Habibi, 2017; Wang et al., 2018; Hou et al., 2019) have exploited a type of 'model-free' RL algorithm known as *Q*-learning (Table 1) to quantitatively learn which cellular behavior, or action, an agent should take on, based on its surroundings (environmental context). In this approach, *state-action* pairs (see above) are mapped to a *reward space* by a quality function, *Q*, which can be roughly viewed as the expectation value of the reward over a series of state-action pairs (i.e., a series of actions and the successive states that they link). *Q*-learning seeks state-action policies which are optimal in the sense of maximizing the overall/cumulative reward. As one might imagine, achieving this goal involves both *exploration* and *exploitation* in the solution space: (i) roughly speaking, '*exploration*' means sampling new, potentially distant regions of a system's universe of possible state-action pairs under the current *policy* (this can be viewed as a long-term/delayed reward), whereas (ii) '*exploitation*' means (re)sampling an already characterized and advantageous region of the space (e.g., a local energy minimum). The *exploration/exploitation* trade-off enters the *Q* equation as the (adjustable) *learning rate*. Intuitively, one can imagine that more exploration occurs relatively early in an RL episode (at which point the solution space, or policy space, has been less mapped-out), whereas the balance might shift towards exploitation in later stages (once the algorithm has learned more productive/rewarding types of actions, corresponding to particular regions of the state-action space).

As an example of the applicability of this type of ML in ABMs, one study developed a 3D hybrid agent-based model of a vascularized tumor, wherein a *Q*-learning algorithm dynamically determined individual cell phenotypes based on features of their surrounding environment (Figure 5), such as local oxygen and glucose concentrations, cell division count, and number of healthy and cancerous neighbors (Zangooei and Habibi, 2017). Comparison with predictions from other, validated ODE-based models (Wodarz and Komarova, 2009; Gerlee, 2013) indicated that the ABM could accurately recapitulate cell phenotype selection and angiogenesis behaviors.

*Q*-learning has also been used to model cell migration behaviors in multicellular systems. Cell migration is an intricate and challenging process to model because a subtle combination of chemotactic gradients, cell···substrate interactions, and other factors influence the direction of movement. One study, which used *Q*-learning to develop cell migration rules in an ABM of *C. elegans* embryogenesis (Wang et al., 2018), trained a deep-*Q* network that optimizes individual cell migratory behaviors in the system. Deep-*Q* networks are a deep-RL approach which integrate deep NNs (e.g., deep convolutional neural nets) with the *Q*-learning framework in order to improve the power and efficiency of a basic RL approach (Alpaydin, 2021); this improvement is achieved by virtue of using a DNN, versus a variant of Bellman's equation from dynamic programming (Eddy, 2004), to represent and optimize the *Q*-function mentioned above (which, again, underlies the mapping of *state-action* pairs and *probabilistic policies* to the *reward* space).  In a similar way, *Q*-learning also has been used to define cell migration behaviors in leader-follower systems (Hou et al., 2019).  In these contexts in particular, RL methods can be seen as a complement to popular 'swarm intelligence'-based approaches (Table 1), such as the particle-swarm, ant-colony, and dragonfly stochastic optimization algorithms (Meraihi et al., 2020; Jin et al., 2021), which feature ants, dragonflies, etc., as agents.

### 3.5  APPLYING ML TO ABM RULE-GENERATION: EXAMPLES, AND EMERGING PRINCIPLES?

Ideally, an ABM's ruleset captures the underlying mechanisms that govern the behaviors of individual entities in response to their local surroundings. Thus, an implication of applying ML to extrapolate agent rules is that the structure and quantitative formulation of the ML model, itself, accurately expresses the decision-making process of an agent, and that the model is generalizable (at least to within some sufficient bounds). As an early example of using ANNs to model agent rules, Gerlee & Anderson (2007) embedded ANNs in cellular agents of a growing tumor to predict cell behaviors, such as proliferation, quiescence, movement, or apoptosis. In that approach, the ANN modeled the "response network" (or rules) governing the behavior of cellular agents in response to their local microenvironments, an assumption being that an NN architecture could reasonably well represent how individual cells enact behaviors in the complex tumor microenvironment. With somewhat similar aims but a different approach, Zangooei & Habibi (2017) applied an RL algorithm to define cancer cell agent behaviors in a growing three-dimensional tumor.  The contrasting approaches used by these studies suggests that it could be interesting to assess their relative strengths in representing cellular decision-making in the intricate microenvironment of a tumor.  Indeed, there is now an opportunity to evaluate how different ML approaches affect the accuracies and generality of ABM predictions.  Augustijn et al. (2020) performed such a head-to-head comparison in an epidemiological ABM of cholera spread: specifically, they contrasted a decision-tree–based algorithm and an EK-driven naïve Bayes approach to simulating decision-making in the ABM.  As might be expected, the study found that the emergent predictions of the ABM varied based on which ML approach was used.  As was emphasized above for ML model validation, we stress here that there is no "one-size-fits-all" approach to integrating ML and ABM. While a key principle is to strive for an optimal 'match' between the structure/architecture of the ML approach and the ABM simulation system, it nevertheless pays to systematically analyze the behavior of as many ML/ABM integration schemes as resources permit.

The choice of ML algorithm likely will be influenced by two factors: (i) the type and availability of data and (ii) the ability to validate the ML algorithm, both on its own and after being integrated with an ABM framework. In epidemiological settings, survey data and EHRs enable the creation of massive training datasets that are amenable to training and validating supervised learning algorithms in isolation to define agent behaviors. Here, several supervised learning algorithms can be trained and the algorithm which highest predictive accuracy can be embedded in an ABM simulation.  In contrast, there is not as much sheer data available describing cell behaviors within tissue contexts. In these situations, training and validating supervised learning algorithms may be less feasible; there, most studies leverage EK-driven ANNs or RL algorithms.  As was alluded to earlier, recall that NNs are "universal predictors," meaning that a functional relationship can be found for any data set.  In biomedical datasets—which are generally high-dimensional and often noisy—this can easily lead to overfitting of the model to the data. To circumvent overfitting, deep learning models require large datasets for validation. Moreover, perturbations to NNs cannot be assessed using only the training data (see the aforementioned notes on causal hierarchy), and require new instances in order to thoroughly examine the effect of perturbations.  We suggest that as many ML/ABM integrations be tested as is feasible, and that justification be provided (again, to the extent possible) for the specific learning algorithm applied for rule-making. Finally, note that an opportunity that can be envisioned in this field is to integrate ANNs with simulations, perturb the ANNs, and study emergent differences in the ABM.

## 4. USING ML TO CALIBRATE MODELS AND REDUCE ABM COMPUTATIONAL COSTS

Typically, ABMs include a variety of parameters that dictate agent behaviors and impact model outcomes. While some of these parameters may be experimentally accessible and well-characterized, such as the time for a cell to divide or the contagious period of an infected individual, often many parameters are unknown and impossible to measure experimentally.  For example, the probability of two cells forming an adherens junction or the physical distance over which a virus spreads from individual to individual are parameters that are difficult to accurately measure.  Also, certain widely-varying parameters may adopt values that are intrinsically quite broadly distributed.  For example, at the molecular level the diffusive properties of proteins and other molecules can vary greatly based on cytosolic crowding, facilitated transport, etc., to such a degree that the distributions of a single parameter (e.g., the translational diffusion coefficient) are quite broadly distributed (greater than an order of magnitude).  An acute challenge in ABM development is calibrating such parameter values so that model outputs are statistically similar to experimentally measured values, including their distributions. Parameter calibration typically involves the modeler formulating an appropriate 'error' or 'fitness' function that compares model outputs with experimental outputs; the calibration algorithm optimizes multiple parameters so as to minimize error or, equivalently, maximize fitness. For example, in an ABM of infectious disease transmission, an error function may be defined as the squared difference between the final fraction of infected individuals in the model versus a real-world example.  The parameter calibration algorithm would then seek an ABM parameter combination that minimizes this error function—a daunting computational task, as exhaustive, brute-force "parameter sweeps" rapidly become intractable, even for relatively coarse sampling, because of a combinatorial explosion in the size of the search space (i.e., the curse of dimensionality (Donoho, 2000)).  ABMs generally have highly multidimensional parameter spaces, making it critical to have an efficient calibration pipeline that can rapidly explore this space and limit the number of parameter combinations that require evaluation.  Genetic algorithms and other "evolutionary algorithms" offer effective stochastic optimization approaches for high-dimensionality searches, as has been recognized in the context of ABMs (Calvez and Hutzler, 2005; Stonedahl and Wilensky, 2011).  Therefore, the following section considers genetic algorithms in a bit more detail, as an example of these types of biologically-inspired ML algorithms and their interplay with ABMs.

### 4.1 GENETIC ALGORITHMS: AN EXAMPLE OF INTEGRATING STOCHASTIC OPTIMIZATION WITH ABMS

Genetic algorithms (GAs) are a widely used ML approach in parameter calibration and, more generally, in any sort of numerical problem that attempts to identify global optima in vast, multi-dimensional search spaces.  Inspired by the native biological processes of molecular evolution and natural selection, as described in a timeless piece by Holland (1992), GAs are particularly adept at locating combinations of parameters (as 'solutions' or 'individuals' in an *in silico* population) that stochastically optimize a fitness function. A description of GAs in terms of the broader landscape of *evolutionary computation* can be found in Foster (2001); more recently, Jin et al. (2021) have provided a pedagogically helpful review of GAs in relation to swarm-based techniques and other population-based 'metaheuristic' approaches for stochastic optimization problems (Table 1).

In general, a GA operates via several distinct stages: (i) *initialization* of a population of individuals as (randomized) parameter combinations that are encoded as chromosomes (e.g., as bit-strings, each chromosome corresponding to one individual), (ii) numerical evaluation of the *fitness* of each individual in the population at cycle $n$, (iii) a *selection step*, wherein parameter combinations/individuals of relatively high fitness are chosen as 'parents' based on specific criteria/protocols, thereby biasing the population towards greater overall fitness (the selection protocol's algorithm and its thresholds can be stochastic to varying degrees, e.g. "tournament selection", "roulette wheel selection" or similar approaches (Zhong et al., 2005)), (iv) the stochastic application of well-defined *genetic operators*, such as crossover (recombination) and mutation, to a subset of the population, thereby yielding the next generation of individuals as 'offspring'. That next, $n+1$th set of individuals then becomes generation $n$, and the steps, from stage (ii) onwards, are iteratively repeated. Over successive iterations, certain allelic variants ('flavors' of a gene) likely become enriched at specific chromosomal regions, indicating convergence with respect to those genes/regions. The GA cycles can terminate after a specific number of iterations/generations, or perhaps once a convergence threshold is reached. At the conclusion of this process, the set of available individuals (with encoded genotypes) will represent various 'solutions' to the original problem—that is, the solution is read-out as the 'genetic sequence' (i.e., genotype) of the final set of chromosomes, representing the 'fittest' individuals (corresponding to optimal phenotypes). As the iterations of *evaluate fitness → select → reproduce/mutate* proceed, with hopeful exploration of new regions of the search space at each stage, the average fitness of a generation approaches more optimal values (e.g., *maximal* traffic flow, *minimal* free energy, *minimal* loss/error function). At that point, the GA can be considered as having converged and identified a parameter combination that optimizes the fitness function.

While GAs can find parameters that optimize multi-objective fitness functions, thus avoiding having to run an ABM for every possible parameter combination, GAs can still be quite computationally expensive. Because of its inherent stochasticity, an ABM must be run multiple times to reach stable values of a single parameter combination (genotype), for a given generation of the iteratively proceeding GA. Thus, as the complexity and computational burden of the ABM increases, traditional GAs become a less computationally feasible option, particularly for calibrating a sophisticated ABM. Nevertheless, note that GAs have been used in tandem with ABMs in areas as diverse as calibrating models of financial and retail markets (Heppenstall et al., 2007; Fabretti, 2013), in parameterizing an ABM "of the functional human brain" (Joyce et al., 2012), and for model refinement and rule-discovery in a 'high-dimensional' ABM of systemic inflammation (Cockrell and An, 2021). An active area of research concerns the development of strategies by which GA/GA-like approaches can navigate a search space in a manner that is more numerically efficient and computationally robust (e.g., to a pathological fitness landscape). Such stochastic optimization methods include, for example, a family of *covariance matrix adaptation–evolution strategy* (CMA-ES) algorithms (see Slowik & Kwasnicka (2020) for this and related approaches) and, somewhat related, *probabilistic model-building GAs* (PMBGAs) that "guide the search for the optimum by building and sampling explicit probabilistic models of promising candidate solutions" (Wikipedia, 2022).

One way to increase the computational efficiency of GAs is to reduce the number of parameters being optimized, thereby reducing the dimensionality of the overall search space of the GA and the number of steps required to achieve convergence. In this context, ML methods can be applied to conduct sensitivity analyses on an ABM and identify the most sensitive/critical parameters to target for calibration. Random forests (RFs), which are composed of an ensemble of decision trees (Table 1), are a popular supervised learning algorithm for conducting sensitivity analysis (Strobl et al., 2007; Strobl et al., 2008; Criminisi et al., 2012). A study by Garg et al. (2019) used RFs to identify sensitive parameters in a multicellular ABM of three different cell populations in vocal fold surgical injury and repair. In that work, the ABM was first run for a variety of input parameter values to generate outputs and create a training dataset that relates input parameter combinations to output values. Then, an RF was trained on this data to classify model outputs based on initial parameter values. The RF hierarchically orders input parameters by Gini index, which is a measure of variance that relates to the probability of incorrectly classifying an output, were the input parameter randomly chosen from the list of all input parameters (the greater the variance, the greater the degree of misclassification). Viewing the Gini index as a measure of *feature importance*, a parameter with a higher Gini value can be seen as more disproportionately influencing the outcome (relative to other parameters), as the model is more likely to produce wrong (misclassified) output if that parameter value was randomly chosen. After training the RF, this study selected the top three parameters associated with each cell type in the model for calibration with a GA (Garg et al., 2019). This integrative and multipronged approach is mentioned here because it reduced the number of parameters required for calibration via the GA, thus improving the computational efficiency of the overall model calibration process.

Beyond computational efficiency, can GAs and ABMs be integrated in ways that might extend the low-level functionality of one (or both) of these fundamental approaches? For example, can GAs enable *adaptable* agents in an ABM? As reviewed by DeAngelis & Diaz (2019), largely in the context of ecological sciences, the plasticity of agent rule-sets and decision-making processes is a key component in achieving greater accuracy and realism in modeling and simulating complex adaptive systems. Here, the decision-making rules and processes that govern the behavior of an individual agent, at a particular time-step in the simulation, are "generally geared to optimize some sort of fitness measure", and—critically—there is the capacity for the decision-making process to change (*evolve*) via selection processes as a simulation unfolds. As concrete examples of approaches that have been taken to incorporate plasticity and heterogeneity in agent behaviors (across individual agent entities, and across time), we note that *fuzzy cognitive maps* (FCMs) have been employed with GAs and agent-based methodologies in at least two distinct ways: (i) In building a framework that uses FCMs to model gene-regulatory networks, Liu et al. (2018) employed a multi-agent GA and random forests to address the high-dimensional search problem that arises in finding optimal parameters for their large FCM-based models. (ii) More recently, Wozniak et al. (Wozniak et al., 2022) devised a GA-based algorithm to efficiently create agent-level FCMs (i.e., one FCM per unique agent, versus a single global FCM for all agents); importantly, the capacity for agent-specific FCMs, or for agents to "have different traits and also follow different rules", enables the emergence of more finely-grained (and more realistic) population-level heterogeneity. GA-based approaches to enable agents to be more adaptable will make ABMs more 'expressive', affording a more realistic and nuanced view of the systems being modeled. Drawing upon the parallel between the 'agents' (real or virtual) in an ABM and those in RL, we note that Sehgal et al. (2019) found that using a GA afforded significantly more efficient discovery of optimal parameterization values for the agent's learning algorithms in a deep RL framework, wherein agents were updated via *Q*-learning approaches (specifically, *deep deterministic policy gradients* [DDPGs] combined with *hindsight experience replay* [HER]); although that work was in the context of robotics, it can be viewed as ABM-related because an RL algorithm's 'agents' are essentially abstracted, intelligent (non-random) agents that enact decisions[6] based upon a host of feedback factors, at the levels of (i) intra-agent (i.e., an agent's internal state), (ii) inter-agent interactions (with neighbors, near or far) and (iii) agent···environment interactions. Finally, we end with one 'inverse' example: Yousefi et al. (2018) took the approach of developing "ensemble metamodels" (in order to reduce the number of models requiring eventual evaluation by a GA) that were trained on ABM-generated data, and then used the metamodels effectively as fitness functions in their GA framework, aimed at solving the constrained optimization problem of on-the-fly resource allocation in hospital emergency departments. We describe this as 'inverse' because, rather than using a GA to aid in an underlying agent-based framework, ABMs were used to inform the GA process (albeit via the ML-based metamodels). We believe that much synergy of this sort is possible.

---

[6] A simple, operational definition of 'decision' can be found in DeAngelis & Diaz (2019) and references therein, who take a decision to be "*wherever one or two (or more) options is/are selected*".

4.2 An Overview of Surrogate Models

Supervised learning algorithms that create a more easily evaluated meta-model or "surrogate model" of an original ABM can also significantly reduce computational burden and make model calibration processes more computationally tractable. As schematized in Figure 6, this approach involves evaluating the ABM on an initial set of parameter combinations by computing the fitness, given an objective function constructed by the modeler. Then, a supervised learning algorithm is trained on this data in order to create a surrogate ML model that can predict ABM outputs for various initial parameter combinations. Finally, parameter-calibration approaches, such as GAs, particle swarm optimization or other methodologies amenable to vast search spaces (Table 1), can be applied to this surrogate model. Often, the surrogate model runs significantly faster than the ABM because the *inference stage* in an ML pipeline involves simply applying the already-trained model to new data (also, the ML model/function is evaluated for single data items instead of an entire simulation worth of data-points). In order to reduce the run-time of an epidemic model, Pereira et al. (2021) employed this general strategy by training a deep neural network (DNN) on data generated by ABM simulations. Application of the DNN (i.e., inference) was more computationally efficient than executing numerous ABM simulations; and, unlike the ABM, the DNN run-time did not increase as the number of ABM agents increased. This DNN-based surrogate model was then used for parameter calibration (Pereira et al., 2021). Other studies have taken similar approaches, for example by using regression algorithms to train surrogate meta-models of an ABM of interest (Tong et al., 2015; Li et al., 2017; Sai et al., 2019; Lutz and Giabbanelli, 2022).

A relatively recent microsimulation study (Cevik et al., 2016), tracking the progression of breast cancer in women, used a novel active learning[7] approach for parameter calibration. In that work, a surrogate ensemble of ANNs (a 'bag' of ANNs, or 'bagANN') was trained based on ABM-generated training datasets. Then, the bagANN model was used to predict fitness for untested parameter combinations. Parameter combinations with low predicted fitness were reevaluated by the ABM, and a refined training dataset was developed to further train the bagANN. The bagANN was repeatedly trained on parameter combinations with increasing fitness, until the fitness converged at some maximal value. In this way, the overall computational pipeline essentially contained an iterative 'bouncing' between the ML (bagANN, in this case) and the ABM stages, not unlike that described by Rand (2006)[8] in a game-theoretic social sciences context. Finally, we note that the biological bagANN study revealed that an active learning approach could find the optimal parameter combination by evaluating only 2% of the parameter space that had been required to be sampled in a prior study devoted to calibrating the same model (Batina et al., 2013; Cevik et al., 2016).

Surrogate ML models also present novel opportunities to capture and explore *continuous* system behaviors via ABMs. ABMs of multicellular interactions can represent cell–cell interactions, such as migration, adhesion and proliferation, that occur over discrete time steps. However, these discrete cell–cell behaviors stem from molecular processes that occur over an effectively *continuous* time domain, such as the rapid expression of proteins driven by complex intracellular signaling cascades. Multiscale models attempt to represent phenomena that span intra-cellular and inter-cellular biological scales in a more realistic and accurate manner than is otherwise possible; this is pursued by employing continuous approaches that predict intracellular signaling dynamics, and using these predictions to update the discrete rules describing cell-cell behaviors. However, a key challenge in these multiscale models is the increased computational cost of evaluating a continuum model, say of reaction kinetics, at each discrete time-step of an ABM. To address such challenges, a newly promising family of approaches views ABMs and conventional, equation-based modeling not as mutually exclusive approaches (Van Dyke Parunak et al., 1998) but rather as opportunities to calibrate ABMs (Ye et al., 2021) and/or 'learn' (in the ML sense and beyond) a system's dynamics in terms of classic, differential-equation–based frameworks; see, e.g. Nardini et al. (2021) for "a promising, novel and unifying approach" for developing and analyzing biological ABMs.

---

[7] Active learning is a form of iterative supervised learning wherein a learning algorithm can ask an information source for the correct labels for unlabeled (input) data; the training is *iterative* in that the querying can occur periodically, and it is *supervised* insofar as the correct labels that inform the learning come from a trusted, reliable source (e.g., a user or other expert, sometimes called an *oracle*).

[8] To our knowledge, this conference paper is one of the earliest documents that recognized a natural 'fit' between the ABM and ML 'cycles'; in it, Rand's proposed "integrated cycle" is a general framework that interleaves ML and ABM methods.

As regards equation-based mechanistic models, a recent study leveraged the training of surrogate models to improve the computational efficiency of a multiscale model of immune cell interactions in the tumor microenvironment (Cess and Finley, 2020). That work utilized an ODE–based mechanistic model to predict macrophage phenotype, based on surrounding cytokine concentrations, and employed an ABM to represent resultant interactions between cells in the tumor. To mitigate the computational burden of evaluating the mechanistic model at each time step, the group trained a NN on the mechanistic model in order to reduce the model to "a simple input/output system", wherein the inputs were local cytokine concentrations and the outputs were cell phenotype. The NN achieved an accuracy exceeding 98%, and—by revealing that the detailed, intracellular mechanistic model could be recapitulated by a simple binary model—reduced the overall computational complexity of the hybrid, multi-scale ABM (Cess and Finley, 2020).

### 4.3 Training Surrogate Models: The Challenges of Extrapolation and of Guiding Principles

While surrogate models present an opportunity to reduce the computational burdens of re-running several ABMs, the choice of supervised learning algorithm, and training process used to obtain a surrogate model, is critical to ensuring that the surrogate model can accurately recapitulate ABM results under many circumstances and conditions. Such conditions include predictions with "out-of-distribution" (OOD) data (Hendrycks and Gimpel, 2017), i.e. those which are highly dissimilar/distant from the data used for training; again, this closely relates to the generalization power of the implemented ML, and recent reviews of OOD challenges can be found in Ghassemi & Fazl-Ersi (2022) and in Sanchez et al. (2022) (see section 4.3 of the latter for a biomedical context). Surrogate models are generally trained on a subset of the parameter space and then applied to predict ABM behaviors in an unexplored region of the parameter space. However, obtaining a well-trained surrogate model that can be successfully applied in a broad range of scenarios (including OOD) simply may be unfeasible for ABMs of 'difficult' systems, such as those which are highly stochastic, which predict a multitude of possibilities for a single combination of parameter values (a one-to-many mapping), which are characterized by exceedingly high intrinsic variability (imagine a difficult functional form, in a high-dimensional space, and with only sparsely-sampled data available), and so on.

## 5. Using ML to Explore an ABM's Parameter Space

After model development and calibration, the behavior of an ABM can be quantitatively explored and used to address various questions. One conceivable approach to such exploration is sensitivity analysis (SA); in this general approach, one perturbs an ABM and its independent parameters, while monitoring predicted changes in dependent variables and other relevant outputs (ten Broeke et al., 2016). SA is an especially fine-grained instrument for probing the behavior of an ABM system. To assess the coarser-scale behavior of a model, e.g. for 'what-if' type analyses, one might create interventions in the ABM (e.g., by setting certain parameter values to specific ranges corresponding to the intervention), make drastic alterations to thresholds, and/or make similarly large-scale changes to the ABM and its input. On a finer scale, examining regions of a parameter space via SA could aim to determine the optimal values (or assess the effects) of an intervention; for example, in an ABM of tumorigenesis the modeler could find the optimal dosage and scheduling of a drug agent to minimize tumor size. Another goal of parameter-space exploration might be to use ABM simulations to predict real-world responses. In an epidemiological ABM of infectious disease transmission, for example, one can use ABM simulations to predict the timeline of disease spread during an ongoing pandemic. Discrete simulation models also have been applied at the level of viral cell-to-cell transmission *within* a single human; for example, using CA approaches to model intra-host HIV-1 spread, Giabbanelli et al. (2019) examined model parameter estimations in terms of prediction accuracies (accounting for available biological/mechanistic information). Recent ABMs have leveraged a variety of unique ML approaches to aid in testing a wide range of perturbations and in characterizing stochastic results (Figure 7). Interestingly, our review of the literature reveals that the goal of parameter-space exploration—whether it be optimizing the efficacy of an intervention, training a predictive model on ABM simulation results, or so on—tends to be closely associated with the *type* of ML algorithm used.

Reinforcement learning methods have been applied to optimization problems in epidemiological and multicellular ABMs, wherein an intervention, such as a drug, is considered an agent and ML is used to find the optimal policy to achieve an output of interest (e.g., minimizing tumor size). A precision medicine, multicellular ABM used deep reinforcement learning (DRL) to find the optimal multi-cytokine therapy dosage for sepsis patients in an ABM of systemic inflammation (Petersen et al., 2019); in that work, the DRL algorithm found the optimal dosage of 12 cytokines to promote patient recovery. RL also has been used, in a multicellular ABM of glioblastoma, to identify an optimal scheduling of Temozolomide treatment for tumor size minimization (Zade et al., 2020). Similarly, RL was also used to optimize radiotherapy treatment of heterogeneous, vascularized tumors (Jalalimanesh et al., 2017a; Jalalimanesh et al., 2017b).

The aforementioned active learning approaches have been used to accelerate the parameter space exploration of ABMs. Using the ABM framework PhysiCell (Ghaffarizadeh et al., 2018), a recent study compared GA-based and active learning–based parameter exploration approaches in the context of tumor and immune cell interactions (Ozik et al., 2019). The goal of the parameter space exploration was to optimize six different immune cell parameters, including apoptosis rate, kill rate and attachment rate, in order to reduce overall tumor cell count. In the GA approach, optimal parameters were found by iteratively selecting parameter combinations that reduced mean tumor cell count. The active learning approach involved training a surrogate RF classifier on the ABM, such that the RF predicts whether a set of input ABM parameters would yield mean tumor cell counts less than a predefined threshold. Then, in a divide-and-conquer–like strategy, the active learning approach selectively samples 'viable' parameter subspaces that yield tumor cell counts less than the threshold in order to find a solution (Ozik et al., 2019). That work illustrates the possibilities of integrating ML-guided adaptive sampling strategies with ABM-based approaches.

A combination of supervised and unsupervised learning methods have been leveraged to use ABM simulations to create predictive models. A study by Nsoesie et al. (2011) evaluated seven different classification algorithms in terms of their abilities to predict the full epidemic curve (i.e., the graphical representation of population-level disease incidence over time, from the first to last infection), given a partial epidemic curve of only the early, middle, or late stages of disease transmission. The study trained these classification algorithms on simulation data from an influenza ABM that modeled the transmission of influenza virus between human agents. Six different variations of the influenza ABM were used to generate a dataset of partial epidemic curves and their corresponding full epidemic curves. Then, supervised learning algorithms were trained to classify partial epidemic curves into one of the six full-epidemic curve categories. The study found that RF classifiers yielded the highest accuracy, whereas linear discriminant analysis had the lowest accuracy (Nsoesie et al., 2011). Another study (Sheikh-Bahaei and Hunt, 2006) used unsupervised fuzzy *c*-means (FCM) classification to predict the biliary transport and excretion properties of new drugs, based on similarities to previously simulated drugs in an ABM of the interactions between *in silico* hepatocyte and drug agents. The study first parametrized biliary transport and excretion properties of experimentally tested drugs in the ABM using a parameter tuning algorithm. Then, the FCM approach was used to cluster and characterize the degree of similarity of a new drug with previously encountered drugs. Based on the degrees of similarity to previously encountered drugs identified by FCM, the biliary transport and excretion properties of the new drug was estimated as a weighted average of that of previously encountered drugs (Sheikh-Bahaei and Hunt, 2006). These examples demonstrate how supervised and unsupervised ML algorithms can be used synergistically with ABMs to predict outcomes for epidemiological and multicellular systems, including from pharmacological perspectives.

Finally, unsupervised ML algorithms have also been used to discover relevant patterns in the results of ABM simulations, such as identifying how differences in single-cell properties in a tumor give rise to distinct tumor spatial organizations. One study (Karolak et al., 2019) co-varied cell radius, cell division age and cell sensitivity to contact inhibition, in a 3D ABM of tumorigenesis, to enable the creation of a simulated library of multicellular tumor organoids. Then, the study ran unsupervised $k$-medians clustering on this library of multicellular tumor organoids to identify four different classes of tumor organoids. The study found that these four classes of organoids respond differently to drug treatment, and identifying which class a real tumor falls into can guide therapy design (Karolak et al., 2019).

## 6. Discussion

### 6.1 Factors Influencing ABM/ML Integration: Data Volume Constraints

ML is a broad term, encompassing a variety of predictive algorithms that each require various levels of data availability to achieve successful learning/training and decision-making. A chief requirement of supervised learning algorithms, for example, is a large amount of accurately-labeled data for purposes of training and testing. Depending on the biological scale of the modeled system, such datasets may or may not be available, thereby affecting the applicability of supervised learning algorithms to ABM. Many epidemiological systems have extensively used survey data and electronic health records to associate patient features (e.g., age, sex, drug use), with disease states, such as diabetic retinopathy or lung cancer stage. Because of the generally high availability of large epidemiological datasets, many epidemiological ABMs have trained supervised learning algorithms on these data to determine agent rules and behaviors in their models (Day et al., 2013; Alexander Jr et al., 2019; Augustijn et al., 2020). In contrast to epidemiological systems, multicellular systems often lack comprehensive data about the detailed mechanisms that drive complex cellular behaviors. In multicellular systems, a range of cellular behaviors, such as migration, proliferation, apoptosis or necrosis, are contingent upon several dynamic and spatiotemporal cues, such as nutrient availability, local cytokine concentrations and intracellular protein expression levels. Current experimental methodologies are challenged to create datasets that associate these spatiotemporal features of the environment with cellular behaviors across the broad populations of heterogeneous cells that comprise various tissues; thus, training supervised learning algorithms to predict cellular behavior based on environmental features is not currently a routine possibility for those sorts of systems. However, several ABMs have augmented the lack of data in this area with either expert-knowledge–driven ML (Gerlee and Anderson, 2007; 2008; Kazmi et al., 2012a; Kazmi et al., 2012b; Al-Mamun et al., 2013; Al-Mamun et al., 2014; Al-Mamun et al., 2016) or RL algorithms (Zangooei and Habibi, 2017; Wang et al., 2018; Hou et al., 2019) to define cellular behaviors within a multicellular ABM. While these recent approaches are not trained on actual experimental datasets, they do enable each cell to autonomously make decisions based on their local environment, thereby realistically modelling cell-to-cell heterogeneity within multicellular systems. These studies illustrate that the limited availability of training data does not necessarily have to be a prohibitively severe constraint in synergistically integrating ML and ABMs.

## 6.2 FACTORS INFLUENCING ABM/ML INTEGRATION: DATA RELEVANCE AND AUGMENTATION

Much effort in ML is devoted to procuring datasets that suffice for training, particularly in deep learning and classification tasks (whether it be for image recognition, speech processing, etc.). While data veracity/accuracy and volume have always been key, as two of the four V's of Big Data (two others being velocity and variety), the importance of a dataset's *relevance* (to the problem at hand) is being increasingly appreciated as part of a recent "data-centric" shift in A.I. (Ng, 2022), and ABMs could play a significant role in that context. As described in Goodfellow et al.'s text (2016), one approach to data augmentation is to simply "*create fake data and add it to the training set*"; for example, for image-related tasks one might obtain synthetic new data from starting images by applying affine transformations, subjecting images to masks or filters, altering intensities, hues, saturation, and so on. In general, ML efforts can leverage an ABM's capacity to efficiently generate large volumes of plausible/reliable simulation data in order to produce new datasets for ML training workflows. As mentioned and cited elsewhere in this Review, this type of synergistic data augmentation or 'sharing' of data (between interleaved ML and ABM workflows) has already been done in the context of ABM/ML integration, and we envision it becoming more commonplace in these fields.

## 6.3 THE TYPE OF ML INFLUENCES PREDICTION ACCURACIES IN INTEGRATED ABM/ML SYSTEMS

Different ML algorithms will generally rely upon fundamentally different computational methods and criteria to make decisions. For example, a logistic regression model relies on a weighted linear combination of predictor variables to classify an object, whereas a naive Bayes model uses conditional probabilities in addressing such tasks (Table 1). Depending on the data, the two different modeling approaches can make slightly divergent predictions from the same datasets. Thus, when integrated with an ABM system, the nature of the different ML algorithms will impact the accuracy of the ABM in simulating real-world systems and the kinds of predictions the ABM generates. For example, two previously discussed studies took alternative approaches in using ML to define the rules that determine human agent water use in an ABM of cholera spread (Abdulkareem et al., 2019; Augustijn et al., 2020). One study trained Bayesian networks to define agent behaviors, while the other used decision trees to derive agent rules of similar form. The ABM using the decision tree approach yielded higher predictions for the number of infected individuals; this discrepancy is unsurprising, as the two ML/ABM approaches differ in the fundamental ways in which each agent behaves, as a consequence of the ML. Considering how the different integrations of ML in these two studies yield different ABM results shows the importance of testing multiple ML integration strategies in order to assess the respective accuracies of the methods.

To circumvent the issue of different ML integration strategies yielding different results, several studies have tested a variety of ML integrations within their ABM prior to deriving model predictions (Sheikh-Bahaei and Hunt, 2006; Nsoesie et al., 2011; Alexander Jr et al., 2019; Gaudou et al., 2020; Hinch et al., 2021). These studies suggest that the computational method underlying an ML algorithm should relate to *how* an (actual) agent could plausibly make decisions in a real-world environment, including, for instance, agent adaptability. For example, whether a cell decides to proliferate, migrate or adhere to its neighbors may be a result of combining weighted inputs from signaling cascades, similar to the linear weighted sum found in logistic regression models. Alternatively, when a human decides whether or not to use a particular water source they may (implicitly) evaluate a series of binary questions, similar to the trajectory through a decision tree. In viewing an integrated ML/ABM system, a key idea is that 'alignment' between the ML approach and the ABM formulation is vital, both for accuracy of the ABM simulations and as regards the issue of whether the computational pipeline's model of the system is limited (e.g., to fitting data, as ML is adept at doing), versus simulating the system with physical realism (a potential benefit of the ABM 'mindset').

## 6.4 INTEGRATED ABM/ML SYSTEMS AND THE IMPORTANCE OF BEING FAIR

The intersection of ML and ABM is a nascent and rapidly-growing field, made possible by the increased publication of peer-reviewed, reproducible and shareable ABM models.  In order to facilitate the continued growth of this field, it is imperative that its members publish models that are validated for given biological contexts, well-documented and maintained, transparent and freely available (open-access, open-source), and otherwise readily usable by all researchers in the many ABM-related communities; though it was conducted six years ago, a careful analysis of over 2000 ABM-related articles found that *"sharing the model code of ABMs is still rare [≈10% of papers] but the practice is now slowly improving"* (Janssen, 2017).  Open-access publication and open-source distribution of validated, clearly-explained and shareable models—which the ABM and related modeling communities can apply, extend, and learn from—is critical in order to advance new techniques and synergistic approaches at the junction of the ML and ABM ecosystems. Indeed, this latter point is essentially a call to apply the "FAIR Principles" of scholarly research (Findability, Accessibility, Interoperability and Reusability (Wilkinson et al., 2016)) to the ABM and ABM/ML fields.

## 7. CONCLUSION

In this Review, we have outlined how several lines of work have incorporated ML in the various stages of developing and deploying ABMs of multicellular or epidemiological systems—from defining individual agent behaviors and optimal rule-sets, to tuning model parameters, to quantitatively exploring a model's sensitivity and features of its parameter space.  Our review of the literature suggests two guiding principles when using ML algorithms in conjunction with ABMs.  First, the *biological scale* of the system (molecular, cellular, organismal, epidemiological, etc.) and the *type of data available* about the system directly impact the type of ML algorithms (supervised, expert knowledge-driven, unsupervised, active, reinforcement learning, etc.) that are applicable in determining and describing agent behaviors, as schematized in Figure 8.  Second, the specific *type of ML algorithm* used with an ABM strongly impacts the emergent results and veracity of predictions by the ABM, making it critical to evaluate multiple ML/ABM integration schemes and select the algorithmic approach with highest accuracy and/or similarity to the decision-making structure of the (actual) modeled system; notably, a similar point was recently made in the social sciences by Brearcliffe & Crooks (2021), who found that *"different ML methods used in the same [ABM] model impact the simulation outcome"*.  Just as there is no universally 'best' approach to ABM or ML individually, there is no one-size-fits-all solution for their integration: systematic explorations of ML/ABM integration schemes would seem to be a worthwhile endeavor.

## ACKNOWLEDGEMENTS

We thank Bill Basener, Phil Bourne and Lei Xie for helpful discussions, support, and reading of the manuscript.  Portions of this work were supported by NSF CAREER award MCB-1350957 and the UVA School of Data Science.  Note that this review provides only a glimpse of ML and ABMs, reflecting our own familiarity with a subset of many potential topics; absence of specific citations reflects only space constraints and not opinions about the significance or merit of earlier work by others.

# REFERENCES

Abdulkareem, S.A., Mustafa, Y.T., Augustijn, E.-W., and Filatova, T. (2019). Bayesian networks for spatial learning: a workflow on using limited survey data for intelligent learning in spatial agent-based models. *GeoInformatica* 23(2)**,** 243-268. doi: 10.1007/s10707-019-00347-0.

Al-Mamun, M., Srisukkham, W., Fall, C., Bass, R., Hossain, A., and Farid, D.M. (2014). "A cellular automaton model for hypoxia effects on tumour growth dynamics", in: *The 8th International Conference on Software, Knowledge, Information Management and Applications (SKIMA 2014)*), 1-8.

Al-Mamun, M.A., Brown, L.J., Hossain, M.A., Fall, C., Wagstaff, L., and Bass, R. (2013). A hybrid computational model for the effects of maspin on cancer cell dynamics. *Journal of Theoretical Biology* 337**,** 150-160. doi: 10.1016/j.jtbi.2013.08.016.

Al-Mamun, M.A., Farid, D.M., Ravenhil, L., Hossain, M.A., Fall, C., and Bass, R. (2016). An in silico model to demonstrate the effects of Maspin on cancer cell dynamics. *Journal of Theoretical Biology* 388**,** 37-49. doi: 10.1016/j.jtbi.2015.10.007.

Alber, M., Buganza Tepole, A., Cannon, W.R., De, S., Dura-Bernal, S., Garikipati, K., et al. (2019). Integrating machine learning and multiscale modeling-perspectives, challenges, and opportunities in the biological, biomedical, and behavioral sciences. *NPJ Digital Medicine* 2**,** 115. doi: 10.1038/s41746-019-0193-y.

Alexander Jr., J., Edwards, R.A., Manca, L., Grugni, R., Bonfanti, G., Emir, B., et al. (2019). Integrating Machine Learning With Microsimulation to Classify Hypothetical, Novel Patients for Predicting Pregabalin Treatment Response Based on Observational and Randomized Data in Patients With Painful Diabetic Peripheral Neuropathy. *Pragmatic and Observational Research* 10**,** 67-76. doi: 10.2147/POR.S214412.

Alpaydin, E. (2021). *Machine Learning.* Cambridge, MA, USA: MIT Press.

An, G. (2006). Concepts for developing a collaborative in silico model of the acute inflammatory response using agent-based modeling. *Journal of Critical Care* 21(1)**,** 105-110. doi: 10.1016/j.jcrc.2005.11.012.

Andreu Perez, J., C. Y. Poon, C., Merrifield, R., and Guang-Zhong, Y. (2015). Big Data for Health. *1208*. doi: 10.1109/jbhi.2015.2450362.

Argüello, R.J., Combes, A.J., Char, R., Gigan, J.-P., Baaziz, A.I., Bousiquot, E., et al. (2020). SCENITH: A Flow Cytometry-Based Method to Functionally Profile Energy Metabolism with Single-Cell Resolution. *Cell Metabolism* 32(6)**,** 1063-1075.e1067. doi: 10.1016/j.cmet.2020.11.007.

Atallah, L., Lo, B., and Yang, G.-Z. (2012). Can pervasive sensing address current challenges in global healthcare? *Journal of Epidemiology and Global Health* 2(1)**,** 1-13. doi: 10.1016/j.jegh.2011.11.005.

Augustijn, E.-W., Abdulkareem, S.A., Sadiq, M.H., and Albabawat, A.A. (2020). "Machine Learning to Derive Complex Behaviour in Agent-Based Modellzing", in: *2020 International Conference on Computer Science and Software Engineering (CSASE)*), 284-289.

Backer, A., Mortele, K., Keulenaer, B., and Parizel, P. (2005). Tuberculosis: Epidemiology, manifestations, and the value of medical imaging in diagnosis. *JBR-BTR : organe de la Société royale belge de radiologie (SRBR) = orgaan van de Koninklijke Belgische Vereniging voor Radiologie (KBVR)* 89**,** 243-250.

Bae, J.W., Paik, E., Kim, K., Singh, K., and Sajjad, M. (2016). Combining Microsimulation and Agent-based Model for Micro-level Population Dynamics. *Procedia Computer Science* 80**,** 507-517. doi: 10.1016/j.procs.2016.05.331.

Bailey, A.M., Thorne, B.C., and Peirce, S.M. (2007). Multi-cell Agent-based Simulation of the Microvasculature to Study the Dynamics of Circulating Inflammatory Cell Trafficking. *Annals of Biomedical Engineering* 35(6)**,** 916-936. doi: 10.1007/s10439-007-9266-1.

Ballas, D., Broomhead, T., and Jones, P.M. (2019). "Spatial Microsimulation and Agent-Based Modelling," in *The Practice of Spatial Analysis: Essays in memory of Professor Pavlos Kanaroglou,* eds. H. Briassoulis, D. Kavroudakis & N. Soulakellis. (Cham: Springer International Publishing), 69-84.

Bareinboim, E., Correa, J.D., Ibeling, D., and Icard, T. (2022). "On Pearl's Hierarchy and the Foundations of Causal Inference," in *Probabilistic and Causal Inference: The Works of Judea Pearl*. Association for Computing Machinery), 507–556.

Batina, N.G., Trentham-Dietz, A., Gangnon, R.E., Sprague, B.L., Rosenberg, M.A., Stout, N.K., et al. (2013). Variation in tumor natural history contributes to racial disparities in breast cancer stage at diagnosis. *Breast cancer research and treatment* 138(2)**,** 519-528. doi: 10.1007/s10549-013-2435-z.

Bhavsar, H., and Ganatra, A. (2012). A Comparative Study of Training Algorithms for Supervised Machine Learning. *International Journal of Soft Computing and Engineering (IJSCE)* 2.

Bonabeau, E. (2002). Agent-based modeling: Methods and techniques for simulating human systems. *Proceedings of the National Academy of Sciences* 99(suppl 3)**,** 7280-7287. doi: 10.1073/pnas.082080899.

Booth, H.P., Khan, O., Fildes, A., Prevost, A.T., Reddy, M., Charlton, J., et al. (2016). Changing Epidemiology of Bariatric Surgery in the UK: Cohort Study Using Primary Care Electronic Health Records. *Obesity Surgery* 26(8), 1900-1905. doi: 10.1007/s11695-015-2032-9.

Bora, Ş., Evren, V., Emek, S., and Çakırlar, I. (2019). Agent-based modeling and simulation of blood vessels in the cardiovascular system. *SIMULATION* 95(4), 297-312. doi: 10.1177/0037549717712602.

Borracci, R.A., and Giorgi, M.A. (2018). Agent-based computational models to explore diffusion of medical innovations among cardiologists. *International Journal of Medical Informatics* 112, 158-165. doi: 10.1016/j.ijmedinf.2018.02.008.

Brearcliffe, D.K., and Crooks, A. (2021). "Creating Intelligent Agents: Combining Agent-Based Modeling with Machine Learning", in: *Proceedings of the 2020 Conference of The Computational Social Science Society of the Americas*, eds. Z. Yang & E. von Briesen: Springer International Publishing), 31-58.

Çağlayan, Ç., Terawaki, H., Chen, Q., Rai, A., Ayer, T., and Flowers, C.R. (2018). Microsimulation Modeling in Oncology. *JCO Clinical Cancer Informatics* (2), 1-11. doi: 10.1200/cci.17.00029.

Calvez, B., and Hutzler, G. (2005). "Parameter Space Exploration of Agent-Based Models", eds. R. Khosla, R.J. Howlett & L.C. Jain (Berlin, Heidelberg: Springer), 633-639.

Casey, J.A., Schwartz, B.S., Stewart, W.F., and Adler, N.E. (2016). Using Electronic Health Records for Population Health Research: A Review of Methods and Applications. *Annual Review of Public Health* 37(1), 61-81. doi: 10.1146/annurev-publhealth-032315-021353.

Cess, C.G., and Finley, S.D. (2020). Multi-scale modeling of macrophage—T cell interactions within the tumor microenvironment. *PLOS Computational Biology* 16(12), e1008519. doi: 10.1371/journal.pcbi.1008519.

Cevik, M., Ali Ergun, M., Stout, N.K., Trentham-Dietz, A., Craven, M., and Alagoz, O. (2016). Using Active Learning for Speeding up Calibration in Simulation Models. *Medical decision making : an international journal of the Society for Medical Decision Making* 36(5), 581-593. doi: 10.1177/0272989x15611359.

Chen, P., Yan, S., Wang, J., Guo, Y., Dong, Y., Feng, X., et al. (2019). Dynamic Microfluidic Cytometry for Single-Cell Cellomics: High-Throughput Probing Single-Cell-Resolution Signaling. *Analytical Chemistry* 91(2), 1619-1626. doi: 10.1021/acs.analchem.8b05179.

Chicco, D. (2017). Ten quick tips for machine learning in computational biology. *BioData Mining* 10(1), 35. doi: 10.1186/s13040-017-0155-3.

Ching, T., Himmelstein, D.S., Beaulieu-Jones, B.K., Kalinin, A.A., Do, B.T., Way, G.P., et al. (2018). Opportunities and obstacles for deep learning in biology and medicine. *Journal of The Royal Society Interface* 15(141), 20170387. doi: 10.1098/rsif.2017.0387.

Chu, S.H., Huang, M., Kelly, R.S., Benedetti, E., Siddiqui, J.K., Zeleznik, O.A., et al. (2019). Integration of Metabolomic and Other Omics Data in Population-Based Study Designs: An Epidemiological Perspective. *Metabolites* 9(6), 117. doi: 10.3390/metabo9060117.

Cockrell, C., and An, G. (2021). Utilizing the Heterogeneity of Clinical Data for Model Refinement and Rule Discovery Through the Application of Genetic Algorithms to Calibrate a High-Dimensional Agent-Based Model of Systemic Inflammation. *Frontiers in Physiology* 12.

Confalonieri, R., Coba, L., Wagner, B., and Besold, T.R. (2021). A historical perspective of explainable Artificial Intelligence. *WIREs Data Mining and Knowledge Discovery* 11(1), e1391. doi: 10.1002/widm.1391.

Cosgrove, J., Butler, J., Alden, K., Read, M., Kumar, V., Cucurull-Sanchez, L., et al. (2015). Agent-Based Modeling in Systems Pharmacology. *CPT: Pharmacometrics & Systems Pharmacology* 4(11), 615-629. doi: 10.1002/psp4.12018.

Criminisi, A., Shotton, J., and Konukoglu, E. (2012). Decision Forests: A Unified Framework for Classification, Regression, Density Estimation, Manifold Learning and Semi-Supervised Learning. *Found. Trends. Comput. Graph. Vis.* 7(2–3), 81–227. doi: 10.1561/0600000035.

Cuevas, E. (2020). An agent-based model to evaluate the COVID-19 transmission risks in facilities. *Computers in Biology and Medicine* 121, 103827. doi: 10.1016/j.compbiomed.2020.103827.

Day, T.E., Ravi, N., Xian, H., and Brugh, A. (2013). An Agent-Based Modeling Template for a Cohort of Veterans with Diabetic Retinopathy. *PLOS ONE* 8(6), e66812. doi: 10.1371/journal.pone.0066812.

DeAngelis, D.L., and Diaz, S.G. (2019). Decision-Making in Agent-Based Modeling: A Current Review and Future Prospectus. *Frontiers in Ecology and Evolution* 6. doi: 10.3389/fevo.2018.00237.

Deutsch, A., Nava-Sedeño, J.M., Syga, S., and Hatzikirou, H. (2021). BIO-LGCA: A cellular automaton modelling class for analysing collective cell migration. *PLOS Computational Biology* 17(6), e1009066. doi: 10.1371/journal.pcbi.1009066.

Donoho, D.L. (2000). "High-dimensional data analysis: The curses and blessings of dimensionality", in: *AMS Conference on Math Challenges of The 21st Century*).

Eddy, S.R. (2004). What is dynamic programming? *Nature biotechnology* 22(7), 909-910. doi: 10.1038/nbt0704-909.

Ehrenstein, V., Nielsen, H., Pedersen, A.B., Johnsen, S.P., and Pedersen, L. (2017). Clinical epidemiology in the era of big data: new opportunities, familiar challenges. *Clinical Epidemiology* 9, 245-250. doi: 10.2147/CLEP.S129779.

Fabretti, A. (2013). On the problem of calibrating an agent based model for financial markets. *Journal of Economic Interaction and Coordination* 8(2), 277-293. doi: 10.1007/s11403-012-0096-3.

Foster, J.A. (2001). Evolutionary Computation. *Nature Reviews: Genetics* 2(6), 428-436. doi: 10.1038/35076523.

Gallaher, J.A., Enriquez-Navas, P.M., Luddy, K.A., Gatenby, R.A., and Anderson, A.R.A. (2018). Spatial Heterogeneity and Evolutionary Dynamics Modulate Time to Recurrence in Continuous and Adaptive Cancer Therapies. *Cancer Research* 78(8), 2127-2139.

Garg, A., Yuen, S., Seekhao, N., Yu, G., Karwowski, J.A.C., Powell, M., et al. (2019). Towards a Physiological Scale of Vocal Fold Agent-Based Models of Surgical Injury and Repair: Sensitivity Analysis, Calibration and Verification. *Applied sciences (Basel, Switzerland)* 9(15). doi: 10.3390/app9152974.

Gaudou, B., Huynh, N.Q., Philippon, D., Brugière, A., Chapuis, K., Taillandier, P., et al. (2020). COMOKIT: A Modeling Kit to Understand, Analyze, and Compare the Impacts of Mitigation Policies Against the COVID-19 Epidemic at the Scale of a City. *Frontiers in Public Health* 8.

Gentleman, R., and Carey, V.J. (2008). "Unsupervised Machine Learning," in *Bioconductor Case Studies,* eds. F. Hahne, W. Huber, R. Gentleman & S. Falcon. (New York, NY: Springer), 137-157.

Gerlee, P. (2013). The model muddle: in search of tumor growth laws. *Cancer Research* 73(8), 2407-2411. doi: 10.1158/0008-5472.can-12-4355.

Gerlee, P., and Anderson, A.R.A. (2007). An evolutionary hybrid cellular automaton model of solid tumour growth. *Journal of Theoretical Biology* 246(4), 583-603. doi: 10.1016/j.jtbi.2007.01.027.

Gerlee, P., and Anderson, A.R.A. (2008). A hybrid cellular automaton model of clonal evolution in cancer: The emergence of the glycolytic phenotype. *Journal of Theoretical Biology* 250(4), 705-722. doi: 10.1016/j.jtbi.2007.10.038.

Ghaffarizadeh, A., Heiland, R., Friedman, S.H., Mumenthaler, S.M., and Macklin, P. (2018). PhysiCell: An open source physics-based cell simulator for 3-D multicellular systems. *PLOS Computational Biology* 14(2), e1005991. doi: 10.1371/journal.pcbi.1005991.

Ghassemi, N., and Fazl-Ersi, E. (2022). A Comprehensive Review of Trends, Applications and Challenges In Out-of-Distribution Detection. *https://arxiv.org/abs/2209.12935*. doi: 10.48550/arXiv.2209.12935.

Giabbanelli, P.J. (2019). "Solving Challenges at the Interface of Simulation and Big Data Using Machine Learning", in: *2019 Winter Simulation Conference (WSC)*), 572-583.

Giabbanelli, P.J., Freeman, C., Devita, J.A., Rosso, N., and Brumme, Z.L. (2019). "Mechanisms for Cell-to-cell and Cell-free Spread of HIV-1 in Cellular Automata Models", in: *Proceedings of the 2019 ACM SIGSIM Conference on Principles of Advanced Discrete Simulation.* (Chicago, IL, USA: Association for Computing Machinery).

Glorot, X., Bordes, A., and Bengio, Y. (2011). "Deep Sparse Rectifier Neural Networks", in: *Proceedings of the Fourteenth International Conference on Artificial Intelligence and Statistics.* (eds.) G. Geoffrey, D. David & D. Miroslav. (Proceedings of Machine Learning Research: PMLR).

Goodfellow, I., Bengio, Y., and Courville, A. (2016). *Deep Learning.* Cambridge, MA, USA: MIT Press.

Gregg, R.W., Shabnam, F., and Shoemaker, J.E. (2021). Agent-based Modeling Reveals Benefits of Heterogeneous and Stochastic Cell Populations During cGAS-mediated IFNβ Production. Bioinformatics 37(10), 1428-1434. doi: 10.1093/bioinformatics/btaa969.

Guillodo, E., Lemey, C., Simonnet, M., Walter, M., Baca-García, E., Masetti, V., et al. (2020). Clinical Applications of Mobile Health Wearable–Based Sleep Monitoring: Systematic Review. *JMIR mHealth and uHealth* 8(4), e10733. doi: 10.2196/10733.

Hastie, T., Tibshirani, R., and Friedman, J.H. (2009). *The elements of statistical learning: Data mining, inference, and prediction.* New York, NY: Springer.

Haykin, S.S. (2009). *Neural Networks and Learning Machines.* New York: Prentice Hall.

Heard, D., Dent, G., Schifeling, T., and Banks, D. (2015). Agent-Based Models and Microsimulation. *Annual Review of Statistics and Its Application* 2(1), 259-272. doi: 10.1146/annurev-statistics-010814-020218.

Hendrycks, D., and Gimpel, K. (2017). "A Baseline for Detecting Misclassified and Out-of-Distribution Examples in Neural Networks".).

Heppenstall, A.J., Evans, A.J., and Birkin, M.H. (2007). Genetic Algorithm Optimisation of An Agent-Based Model for Simulating a Retail Market. *Environment and Planning B: Planning and Design* 34(6), 1051-1070. doi: 10.1068/b32068.

Hinch, R., Probert, W.J.M., Nurtay, A., Kendall, M., Wymant, C., Hall, M., et al. (2021). OpenABM-Covid19—An agent-based model for non-pharmaceutical interventions against COVID-19 including contact tracing. *PLOS Computational Biology* 17(7), e1009146. doi: 10.1371/journal.pcbi.1009146.

Holland, J.H. (1992). Genetic Algorithms. *Scientific American* 267(1), 66-73.

Hou, H., Gan, T., Yang, Y., Zhu, X., Liu, S., Guo, W., et al. (2019). Using deep reinforcement learning to speed up collective cell migration. *BMC Bioinformatics* 20(Suppl 18). doi: 10.1186/s12859-019-3126-5.

Hunt, C.A., Kennedy, R.C., Kim, S.H.J., and Ropella, G.E.P. (2013). Agent-based modeling: a systematic assessment of uses cases and requirements for enhancing pharmaceutical research and development productivity. *WIREs Systems Biology and Medicine* 5(4), 461-480. doi: 10.1002/wsbm.1222.

Hwang, B., Lee, J.H., and Bang, D. (2018). Single-cell RNA sequencing technologies and bioinformatics pipelines. *Experimental & Molecular Medicine* 50(8), 1-14. doi: 10.1038/s12276-018-0071-8.

Irish, J.M., Kotecha, N., and Nolan, G.P. (2006). Mapping normal and cancer cell signalling networks: towards single-cell proteomics. *Nature Reviews Cancer* 6(2), 146-155. doi: 10.1038/nrc1804.

Jalalimanesh, A., Haghighi, H.S., Ahmadi, A., Hejazian, H., and Soltani, M. (2017a). Multi-objective optimization of radiotherapy: distributed Q-learning and agent-based simulation. *Journal of Experimental & Theoretical Artificial Intelligence* 29(5), 1071-1086. doi: 10.1080/0952813x.2017.1292319.

Jalalimanesh, A., Shahabi Haghighi, H., Ahmadi, A., and Soltani, M. (2017b). Simulation-based optimization of radiotherapy: Agent-based modeling and reinforcement learning. *Mathematics and Computers in Simulation* 133, 235-248. doi: https://doi.org/10.1016/j.matcom.2016.05.008.

Janssen, M.A. (2017). The Practice of Archiving Model Code of Agent-Based Models. *Journal of Artificial Societies and Social Simulation* 20(1), 2. doi: 10.18564/jasss.3317.

Ji, Z., Su, J., Wu, D., Peng, H., Zhao, W., Zhao, B.N., et al. (2016). Predicting the impact of combined therapies on myeloma cell growth using a hybrid multi-scale agent-based model. *Oncotarget* 8(5), 7647-7665. doi: 10.18632/oncotarget.13831.

Jiménez-Luna, J., Grisoni, F., and Schneider, G. (2020). Drug discovery with explainable artificial intelligence. *Nature Machine Intelligence* 2(10), 573-584. doi: 10.1038/s42256-020-00236-4.

Jin, Y., Wang, H., and Sun, C. (2021). "Evolutionary and Swarm Optimization," in *Data-Driven Evolutionary Optimization: Integrating Evolutionary Computation, Machine Learning and Data Science,* eds. Y. Jin, H. Wang & C. Sun. (Cham: Springer International Publishing), 53-101.

Jones, D.T. (2019). Setting the standards for machine learning in biology. *Nature Reviews Molecular Cell Biology* 20(11), 659-660. doi: 10.1038/s41580-019-0176-5.

Joyce, K.E., Hayasaka, S., and Laurienti, P.J. (2012). A GENETIC ALGORITHM FOR CONTROLLING AN AGENT-BASED MODEL OF THE FUNCTIONAL HUMAN BRAIN. *Biomedical sciences instrumentation* 48, 210-217.

Kaelbling, L.P., Littman, M.L., and Moore, A.W. (1996). Reinforcement Learning: A Survey. *Journal of Artificial Intelligence Research* 4, 237-285. doi: 10.1613/jair.301.

Karolak, A., Poonja, S., and Rejniak, K.A. (2019). Morphophenotypic classification of tumor organoids as an indicator of drug exposure and penetration potential. *PLOS Computational Biology* 15(7), e1007214. doi: 10.1371/journal.pcbi.1007214.

Kassambara, A. (2017). *Practical Guide to Cluster Analysis in R: Unsupervised Machine Learning.* STHDA.

Kazmi, N., Hossain, M.A., and Phillips, R.M. (2012a). A hybrid cellular automaton model of solid tumor growth and bioreductive drug transport. *IEEE/ACM transactions on computational biology and bioinformatics* 9(6), 1595-1606. doi: 10.1109/tcbb.2012.118.

Kazmi, N., Hossain, M.A., Phillips, R.M., Al-Mamun, M.A., and Bass, R. (2012b). Avascular tumour growth dynamics and the constraints of protein binding for drug transportation. *Journal of Theoretical Biology* 313, 142-152. doi: 10.1016/j.jtbi.2012.07.026.

Kim, Y., Powathil, G., Kang, H., Trucu, D., Kim, H., Lawler, S., et al. (2015). Strategies of Eradicating Glioma Cells: A Multi-Scale Mathematical Model with MiR-451-AMPK-mTOR Control. *PLOS ONE* 10(1), e0114370. doi: 10.1371/journal.pone.0114370.

Krieger, N. (2003). Place, Space, and Health: GIS and Epidemiology. *Epidemiology* 14(4), 384-385. doi: 10.1097/01.ede.0000071473.69307.8a.

Kulkarni, A., Anderson, A.G., Merullo, D.P., and Konopka, G. (2019). Beyond bulk: a review of single cell transcriptomics methodologies and applications. *Current Opinion in Biotechnology* 58, 129-136. doi: 10.1016/j.copbio.2019.03.001.

Kulkarni, P. (2012). *Reinforcement and Systemic Machine Learning for Decision Making.* John Wiley & Sons.

LeCun, Y., Bengio, Y., and Hinton, G. (2015). Deep learning. *Nature* 521(7553), 436-444. doi: 10.1038/nature14539.

Li, T., Cheng, Z., and Zhang, L. (2017). Developing a Novel Parameter Estimation Method for Agent-Based Model in Immune System Simulation under the Framework of History Matching: A Case Study on Influenza A Virus Infection. *International Journal of Molecular Sciences* 18(12). doi: 10.3390/ijms18122592.

Ligmann-Zielinska, A., Siebers, P.-O., Magliocca, N., Parker, D.C., Grimm, V., Du, J., et al. (2020). 'One Size Does Not Fit All': A Roadmap of Purpose-Driven Mixed-Method Pathways for Sensitivity Analysis of Agent-Based Models. *Journal of Artificial Societies and Social Simulation* 23(1), 6.

Lindau, S.T., Makelarski, J.A., Kaligotla, C., Abramsohn, E.M., Beiser, D.G., Chou, C., et al. (2021). Building and experimenting with an agent-based model to study the population-level impact of CommunityRx, a clinic-based community resource referral intervention. *PLOS Computational Biology* 17(10), e1009471. doi: 10.1371/journal.pcbi.1009471.

Liu, L., and Liu, J. (2018). Inferring gene regulatory networks with hybrid of multi-agent genetic algorithm and random forests based on fuzzy cognitive maps. *Applied Soft Computing* 69, 585-598. doi: https://doi.org/10.1016/j.asoc.2018.05.009.

Longo, D., Peirce, S.M., Skalak, T.C., Davidson, L., Marsden, M., Dzamba, B., et al. (2004). Multicellular computer simulation of morphogenesis: blastocoel roof thinning and matrix assembly in Xenopus laevis. *Developmental Biology* 271(1), 210-222. doi: 10.1016/j.ydbio.2004.03.021.

Lutz, C.B., and Giabbanelli, P.J. (2022). When Do We Need Massive Computations to Perform Detailed COVID-19 Simulations? *Advanced Theory and Simulations* 5(2), 2100343. doi: 10.1002/adts.202100343.

Marshall, B.D.L., and Galea, S. (2015). Formalizing the Role of Agent-Based Modeling in Causal Inference and Epidemiology. *American Journal of Epidemiology* 181(2), 92-99. doi: 10.1093/aje/kwu274.

Martin, K.S., Blemker, S.S., and Peirce, S.M. (2015). Agent-based computational model investigates muscle-specific responses to disuse-induced atrophy. *Journal of Applied Physiology* 118(10), 1299-1309. doi: 10.1152/japplphysiol.01150.2014.

Marx, V. (2019). A dream of single-cell proteomics. *Nature Methods* 16(9), 809-812. doi: 10.1038/s41592-019-0540-6.

Maturana, E.L.d., Pineda, S., Brand, A., Steen, K.V., and Malats, N. (2016). Toward the integration of Omics data in epidemiological studies: still a "long and winding road". *Genetic Epidemiology* 40(7), 558-569. doi: 10.1002/gepi.21992.

Meraihi, Y., Ramdane-Cherif, A., Acheli, D., and Mahseur, M. (2020). Dragonfly algorithm: A comprehensive review and applications. *Neural Computing and Applications* 32(21), 16625-16646. doi: 10.1007/s00521-020-04866-y.

Mitchell, T.M. (1997). *Machine Learning.* New York: McGraw-Hill.

Nardini, J.T., Baker, R.E., Simpson, M.J., and Flores, K.B. (2021). Learning differential equation models from stochastic agent-based model simulations. *Journal of the Royal Society, Interface* 18(176), 20200987. doi: 10.1098/rsif.2020.0987.

Ng, A. (2022). *Unbiggen AI: Farewell, Big Data* [Online]. Available: https://spectrum.ieee.org/andrew-ng-data-centric-ai [Accessed].

Nicholson, D.N., and Greene, C.S. (2020). Constructing knowledge graphs and their biomedical applications. *Computational and Structural Biotechnology Journal* 18, 1414-1428. doi: 10.1016/j.csbj.2020.05.017.

Nsoesie, E.O., Beckman, R., Marathe, M., and Lewis, B. (2011). Prediction of an Epidemic Curve: A Supervised Classification Approach. *Statistical communications in infectious diseases* 3(1). doi: 10.2202/1948-4690.1038.

Oduola, W.O., and Li, X. (2018). Multiscale Tumor Modeling With Drug Pharmacokinetic and Pharmacodynamic Profile Using Stochastic Hybrid System. *Cancer Informatics* 17, 1176935118790262. doi: 10.1177/1176935118790262.

Ozik, J., Collier, N., Heiland, R., An, G., and Macklin, P. (2019). Learning-accelerated discovery of immune-tumour interactions. *Molecular Systems Design & Engineering* 4(4), 747-760. doi: 10.1039/c9me00036d.

Park, C., Took, C.C., and Seong, J.-K. (2018). Machine learning in biomedical engineering. *Biomedical Engineering Letters* 8(1), 1-3. doi: 10.1007/s13534-018-0058-3.

Park, Y., and Sylla, I. (2021). Agent-based Modeling to Evaluate Nosocomial COVID-19 Infections and Related Policies. 7.

Pearl, J. (2000). *Causality: Models, Reasoning, and Inference.* Cambridge, U.K. ; New York: Cambridge University Press.

Peirce, S.M., Van Gieson, E.J., and Skalak, T.C. (2004). Multicellular simulation predicts microvascular patterning and in silico tissue assembly. *The FASEB Journal* 18(6), 731-733. doi: 10.1096/fj.03-0933fje.

Peng, G.C.Y., Alber, M., Buganza Tepole, A., Cannon, W.R., De, S., Dura-Bernal, S., et al. (2021). Multiscale Modeling Meets Machine Learning: What Can We Learn? *Archives of Computational Methods in Engineering* 28(3), 1017-1037. doi: 10.1007/s11831-020-09405-5.

Pereira, F.H., Schimit, P.H.T., and Bezerra, F.E. (2021). A deep learning based surrogate model for the parameter identification problem in probabilistic cellular automaton epidemic models. *Computer Methods and Programs in Biomedicine* 205, 106078. doi: 10.1016/j.cmpb.2021.106078.

Perez-Pozuelo, I., Spathis, D., Clifton, E.A.D., and Mascolo, C. (2021). "Chapter 3 - Wearables, smartphones, and artificial intelligence for digital phenotyping and health," in *Digital Health,* eds. S. Syed-Abdul, X. Zhu & L. Fernandez-Luque. Elsevier), 33-54.

Petersen, B.K., Yang, J., Grathwohl, W.S., Cockrell, C., Santiago, C., An, G., et al. (2019). Deep Reinforcement Learning and Simulation as a Path Toward Precision Medicine. *Journal of Computational Biology* 26(6), 597-604. doi: 10.1089/cmb.2018.0168.

Ponjoan, A., Garre-Olmo, J., Blanch, J., Fages, E., Alves-Cabratosa, L., Martí-Lluch, R., et al. (2019). Epidemiology of dementia: prevalence and incidence estimates using validated electronic health records from primary care. *Clinical Epidemiology* 11, 217-228. doi: 10.2147/clep.s186590.

Potter, S.S. (2018). Single-cell RNA sequencing for the study of development, physiology and disease. *Nature Reviews Nephrology* 14(8), 479-492. doi: 10.1038/s41581-018-0021-7.

Preim, B., Klemm, P., Hauser, H., Hegenscheid, K., Oeltze, S., Toennies, K., et al. (2016). "Visual Analytics of Image-Centric Cohort Studies in Epidemiology", eds. L. Linsen, B. Hamann & H.-C. Hege (Cham: Springer International Publishing), 221-248.

Puterman, M.L. (1990). "Chapter 8 Markov decision processes," in *Handbooks in Operations Research and Management Science*. Elsevier), 331-434.

Rand, W. (2006). "Machine Learning Meets Agent-based Modeling: When Not to Go to a Bar", in: *Conference on Social Agents: Results and Prospects*).

Raschka, S. (2018). Model Evaluation, Model Selection, and Algorithm Selection in Machine Learning. *ArXiv* abs/1811.12808.

Robertson, S.H., Smith, C.K., Langhans, A.L., McLinden, S.E., Oberhardt, M.A., Jakab, K.R., et al. (2007). Multiscale computational analysis of Xenopus laevis morphogenesis reveals key insights of systems-level behavior. *BMC systems biology* 1, 46. doi: 10.1186/1752-0509-1-46.

Rockett, R.J., Arnott, A., Lam, C., Sadsad, R., Timms, V., Gray, K.-A., et al. (2020). Revealing COVID-19 transmission in Australia by SARS-CoV-2 genome sequencing and agent-based modeling. *Nature Medicine* 26(9), 1398-1404. doi: 10.1038/s41591-020-1000-7.

Rytkönen, M.J.P. (2004). Not all maps are equal: GIS and spatial analysis in epidemiology. *International Journal of Circumpolar Health* 63(1), 9-24. doi: 10.3402/ijch.v63i1.17542.

Sai, A., Vivas-Valencia, C., Imperiale, T.F., and Kong, N. (2019). Multiobjective Calibration of Disease Simulation Models using Gaussian Processes. *Medical decision making : an international journal of the Society for Medical Decision Making* 39(5), 540-552. doi: 10.1177/0272989x19862560.

Sanchez, P., Voisey, J.P., Xia, T., Watson, H.I., O'Neil, A.Q., and Tsaftaris, S.A. (2022). Causal Machine Learning for Healthcare and Precision Medicine. *Royal Society Open Science* 9(8), 220638. doi: 10.1098/rsos.220638.

Saracci, R. (2018). Epidemiology in wonderland: Big Data and precision medicine. *European Journal of Epidemiology* 33(3), 245-257. doi: 10.1007/s10654-018-0385-9.

Sehgal, A., La, H., Louis, S., and Nguyen, H. (2019). "Deep Reinforcement Learning Using Genetic Algorithm for Parameter Optimization", in: *2019 Third IEEE International Conference on Robotic Computing (IRC)*), 596-601.

Serra, A., Galdi, P., and Tagliaferri, R. (2018). Machine learning for bioinformatics and neuroimaging. *WIREs Data Mining and Knowledge Discovery* 8(5), e1248. doi: 10.1002/widm.1248.

Sheikh-Bahaei, S., and Hunt, C.A. (2006). "Prediction of in Vitro Hepatic Biliary Excretion using Stochastic Agent-Based Modeling and Fuzzy Clustering", in: *Proceedings of the 2006 Winter Simulation Conference*), 1617-1624.

Shoukat, A., and Moghadas, S.M. (2020). Agent-Based Modelling: An Overview with Application to Disease Dynamics. *ArXiv* abs/2007.04192.

Singh, A., Thakur, N., and Sharma, A. (2016). "A review of supervised machine learning algorithms", in: *2016 3rd International Conference on Computing for Sustainable Global Development (INDIACom)*), 1310-1315.

Slowik, A., and Kwasnicka, H. (2020). Evolutionary Algorithms and Their Applications to Engineering Problems. *Neural Computing and Applications* 32(16)**,** 12363-12379. doi: 10.1007/s00521-020-04832-8.

Smith-Bindman, R., Miglioretti, D.L., and Larson, E.B. (2008). Rising Use Of Diagnostic Medical Imaging In A Large Integrated Health System. *Health Affairs* 27(6)**,** 1491-1502. doi: 10.1377/hlthaff.27.6.1491.

Soheilypour, M., and Mofrad, M.R.K. (2018). Agent-Based Modeling in Molecular Systems Biology. *BioEssays: News and reviews in molecular, cellular and developmental biology* 40(7), e1800020. doi: 10.1002/bies.201800020.

Stonedahl, F., and Wilensky, U. (2011). "Finding Forms of Flocking: Evolutionary Search in ABM Parameter-Spaces", eds. T. Bosse, A. Geller & C.M. Jonker (Berlin, Heidelberg: Springer), 61-75.

Strobl, C., Boulesteix, A.-L., Kneib, T., Augustin, T., and Zeileis, A. (2008). Conditional variable importance for random forests. *BMC bioinformatics* 9**,** 307. doi: 10.1186/1471-2105-9-307.

Strobl, C., Boulesteix, A.-L., Zeileis, A., and Hothorn, T. (2007). Bias in random forest variable importance measures: illustrations, sources and a solution. *BMC bioinformatics* 8**,** 25. doi: 10.1186/1471-2105-8-25.

Su, C., Tong, J., Zhu, Y., Cui, P., and Wang, F. (2020). Network embedding in biomedical data science. *Briefings in Bioinformatics* 21(1)**,** 182-197. doi: 10.1093/bib/bby117.

Sutton, R.S. (1992). "Introduction: The Challenge of Reinforcement Learning," in *Reinforcement Learning,* ed. R.S. Sutton.  (Boston, MA: Springer US), 1-3.

Tang, B., Pan, Z., Yin, K., and Khateeb, A. (2019). Recent Advances of Deep Learning in Bioinformatics and Computational Biology. *Frontiers in Genetics* 10**,** 214. doi: 10.3389/fgene.2019.00214.

Taylor, H.B., Khuong, A., Wu, Z., Xu, Q., Morley, R., Gregory, L., et al. (2017). Cell segregation and border sharpening by Eph receptor-ephrin-mediated heterotypic repulsion. *J R Soc Interface* 14(132). doi: 10.1098/rsif.2017.0338.

Tchito Tchapga, C., Mih, T.A., Tchagna Kouanou, A., Fozin Fonzin, T., Kuetche Fogang, P., Mezatio, B.A., et al. (2021). Biomedical Image Classification in a Big Data Architecture Using Machine Learning Algorithms. *Journal of Healthcare Engineering* 2021**,** e9998819. doi: 10.1155/2021/9998819.

ten Broeke, G., van Voorn, G., and Ligtenberg, A. (2016). Which Sensitivity Analysis Method Should I Use for My Agent-Based Model? *Journal of Artificial Societies and Social Simulation* 19(1)**,** 5.

ten Broeke, G., Voorn, G.v., and Ligtenberg, A. (2014). "Sensitivity analysis for agent-based models: a low complexity test-case", in: *Social Simulation Conference (SSC'14), Barcelona, Spain*).

Thomas, D.C. (2006). High-Volume "-Omics" Technologies and the Future of Molecular Epidemiology. *Epidemiology* 17(5)**,** 490-491. doi: 10.1097/01.ede.0000229950.29674.68.

Thorne, B., Hayenga, H., Humphrey, J., and Peirce, S. (2011). Toward a Multi-Scale Computational Model of Arterial Adaptation in Hypertension: Verification of a Multi-Cell Agent Based Model. *Frontiers in Physiology* 2.

Thorne, B.C., Bailey, A.M., DeSimone, D.W., and Peirce, S.M. (2007a). Agent-based modeling of multicell morphogenic processes during development. *Birth Defects Res C Embryo Today* 81(4)**,** 344-353. doi: 10.1002/bdrc.20106.

Thorne, B.C., Bailey, A.M., and Peirce, S.M. (2007b). Combining experiments with multi-cell agent-based modeling to study biological tissue patterning. *Brief Bioinform* 8(4)**,** 245-257. doi: 10.1093/bib/bbm024.

Tong, X., Chen, J., Miao, H., Li, T., and Zhang, L. (2015). Development of an Agent-Based Model (ABM) to Simulate the Immune System and Integration of a Regression Method to Estimate the Key ABM Parameters by Fitting the Experimental Data. *PLoS ONE* 10(11). doi: 10.1371/journal.pone.0141295.

Tracy, M., Cerdá, M., and Keyes, K.M. (2018). Agent-Based Modeling in Public Health: Current Applications and Future Directions. *Annual Review of Public Health* 39(1)**,** 77-94. doi: 10.1146/annurev-publhealth-040617-014317.

Van Dyke Parunak, H., Savit, R., and Riolo, R.L. (1998). "Agent-Based Modeling vs. Equation-Based Modeling: A Case Study and Users' Guide", in: *Multi-Agent Systems and Agent-Based Simulation*, eds. J.S. Sichman, R. Conte & N. Gilbert: Springer Berlin Heidelberg), 10-25.

Vilone, G., and Longo, L. (2021). Notions of explainability and evaluation approaches for explainable artificial intelligence. *Information Fusion* 76**,** 89-106. doi: 10.1016/j.inffus.2021.05.009.

Virgilio, K.M., Martin, K.S., Peirce, S.M., and Blemker, S.S. (2018). Agent-based model illustrates the role of the microenvironment in regeneration in healthy and mdx skeletal muscle. *Journal of Applied Physiology* 125(5)**,** 1424-1439. doi: 10.1152/japplphysiol.00379.2018.

von Rueden, L., Mayer, S., Beckh, K., Georgiev, B., Giesselbach, S., Heese, R., et al. (2021). Informed Machine Learning - A Taxonomy and Survey of Integrating Prior Knowledge into Learning Systems. *IEEE Transactions on Knowledge and Data Engineering***,** 1-1. doi: 10.1109/tkde.2021.3079836.

Walpole, J., Chappell, J.C., Cluceru, J.G., Mac Gabhann, F., Bautch, V.L., and Peirce, S.M. (2015). Agent-based model of angiogenesis simulates capillary sprout initiation in multicellular networks. *Integrative Biology: Quantitative Biosciences from Nano to Macro* 7(9), 987-997. doi: 10.1039/c5ib00024f.

Walpole, J., Gabhann, F.M., Peirce, S.M., and Chappell, J.C. (2017). Agent-based computational model of retinal angiogenesis simulates microvascular network morphology as a function of pericyte coverage. *Microcirculation* 24(8), e12393. doi: 10.1111/micc.12393.

Wang, Z., Wang, D., Li, C., Xu, Y., Li, H., and Bao, Z. (2018). Deep reinforcement learning of cell movement in the early stage of C.elegans embryogenesis. *Bioinformatics* 34(18), 3169-3177. doi: 10.1093/bioinformatics/bty323.

Wang, Z., Zhang, L., Sagotsky, J., and Deisboeck, T.S. (2007). Simulating non-small cell lung cancer with a multiscale agent-based model. *Theoretical Biology and Medical Modelling* 4(1), 50. doi: 10.1186/1742-4682-4-50.

Warner, V.H., Sivakumar, N., Peirce, M. S., Lazzara, J. M. (2019). Multiscale computational models of cancer. *Current Opinion in Biomedical Engineering* 11. doi: https://doi.org/10.1016/j.cobme.2019.11.002.

Wikipedia, c. (2022). "Estimation of distribution algorithm".).

Wilkinson, M.D., Dumontier, M., Aalbersberg, I.J., Appleton, G., Axton, M., Baak, A., et al. (2016). The FAIR Guiding Principles for scientific data management and stewardship. *Scientific data* 3, 160018. doi: 10.1038/sdata.2016.18.

Willett, D.S., Stelinski, L.L., and Lapointe, S.L. (2015). Using response surface methods to explore and optimize mating disruption of the leafminer Phyllocnistis citrella (Lepidoptera: Gracillariidae). *Frontiers in Ecology and Evolution* 3, 30. doi: 10.3389/fevo.2015.00030.

Wodarz, D., and Komarova, N. (2009). Towards predictive computational models of oncolytic virus therapy: basis for experimental validation and model selection. *PLOS ONE* 4(1), e4271. doi: 10.1371/journal.pone.0004271.

Woelke, A.L., von Eichborn, J., Murgueitio, M.S., Worth, C.L., Castiglione, F., and Preissner, R. (2011). Development of immune-specific interaction potentials and their application in the multi-agent-system VaccImm. *PLOS ONE* 6(8), e23257. doi: 10.1371/journal.pone.0023257.

Wozniak, M.K., Mkhitaryan, S., and Giabbanelli, P.J. (2022). "Automatic Generation of Individual Fuzzy Cognitive Maps from Longitudinal Data", in: *Computational Science – ICCS 2022: 22nd International Conference, London, UK, June 21–23, 2022, Proceedings, Part III.* (London, United Kingdom: Springer-Verlag).

Wu, M., Perroud, T.D., Srivastava, N., Branda, C.S., Sale, K.L., Carson, B.D., et al. (2012). Microfluidically-unified cell culture, sample preparation, imaging and flow cytometry for measurement of cell signaling pathways with single cell resolution. *Lab on a Chip* 12(16), 2823-2831. doi: 10.1039/c2lc40344g.

Wu, M., Piccini, M., Koh, C.-Y., Lam, K.S., and Singh, A.K. (2013). Single Cell MicroRNA Analysis Using Microfluidic Flow Cytometry. *PLOS ONE* 8(1), e55044. doi: 10.1371/journal.pone.0055044.

Xu, Q., Ozturk, M.C., and Cinar, A. (2018). Agent-Based Modeling of Immune Response to Study the Effects of Regulatory T Cells in Type 1 Diabetes. *Processes* 6(9), 141. doi: 10.3390/pr6090141.

Ye, P., Chen, Y., Zhu, F., Lv, Y., Lu, W., and Wang, F.Y. (2021). Bridging the Micro and Macro: Calibration of Agent-Based Model Using Mean-Field Dynamics. *IEEE Transactions on Cybernetics*, 1-10. doi: 10.1109/tcyb.2021.3089712.

Yousefi, M., Ferreira, R.P.M., Kim, J.H., and Fogliatto, F.S. (2018). Chaotic genetic algorithm and Adaboost ensemble metamodeling approach for optimum resource planning in emergency departments. *Artificial Intelligence in Medicine* 84, 23-33. doi: 10.1016/j.artmed.2017.10.002.

Zade, E.A., Haghighi, S.S., and Soltani, M. (2020). Reinforcement learning for optimal scheduling of Glioblastoma treatment with Temozolomide. *Computer Methods and Programs in Biomedicine* 193, 105443. doi: 10.1016/j.cmpb.2020.105443.

Zangooei, M.H., and Habibi, J. (2017). Hybrid multiscale modeling and prediction of cancer cell behavior. *PLOS ONE* 12(8), e0183810. doi: 10.1371/journal.pone.0183810.

Zhang, W., Valencia, A., and Chang, N.B. (2021). Synergistic Integration Between Machine Learning and Agent-Based Modeling: A Multidisciplinary Review. *IEEE transactions on neural networks and learning systems* PP. doi: 10.1109/TNNLS.2021.3106777.

Zhong, J., Hu, X., Zhang, J., and Gu, M. (2005). "Comparison of Performance between Different Selection Strategies on Simple Genetic Algorithms", in: *International Conference on Computational Intelligence for Modelling, Control and Automation and International Conference on Intelligent Agents, Web Technologies and Internet Commerce (CIMCA-IAWTIC'06)*), 1115-1121.

# TABLES

**Table 1**. **This taxonomy of ML algorithms** organizes a few popular approaches using a scheme based on Mitchell's definition of ML. Specifically, we list (i) the type of *Task* each algorithm can address, (ii) how the algorithm measures the *Performance* with respect to that task (an "objective function" typically quantifies accuracy), and (iii) the learning process, or *Experience*, by which an algorithm tunes parameters (generally by optimizing the *Performance* function) in order to improve its accuracy for a given task (estimation, classification, etc.). This Table is meant to be viewed flexibly. For instance, though often associated with supervised learning, NNs span multiple types of ML; as an example, autoencoders (AEs) are often built via NNs to learn optimal representations/models from unlabeled data by minimizing a "reconstruction error" for generating original/input data from a compressed (latent space) representation of that starting information, and in that way AEs can be viewed as a general form of unsupervised learning.

| Type of ML (category, or "learning style") | Sample algorithms | Task | Performance | Experience |
|---|---|---|---|---|
| **Supervised, including semi-supervised or "expert-knowledge driven"** (i.e., at least some labeled data are used) | Linear Regression | Predict a continuous outcome from input parameters/features (numerical estimation) | Sum of squared errors | Iteratively, via gradient descent |
| | Logistic Regression | Predict discrete outcomes (e.g., binary) from input data (classification task) | Logarithmic loss (or 'cross-entropy loss') function | Iteratively, via gradient descent |
| | Naïve Bayes | Predict data labels based on naive prior probability distributions (assumes independent features) | Negative joint likelihood function | For a given problem instance, the class label yielding the largest probability (i.e., a maximum *a posteriori* [MAP] decision rule) |
| | Decision Trees | Predict the sequence of predictor variables that classify a sample within a particular category (vs others) | Gini index (relates to the *relative mean absolute difference*); seek a tree that accounts for most of the data, without excessive number of levels | Can alter number of levels, node-splitting functions (e.g., based on Kullback-Leibler divergence), can try ensemble methods (e.g. random forests) |
| | Support Vector Machine (SVM) | Computes hyperplane that optimally separates points in a dataset, e.g. for classification or regression tasks | Maximize *margin* of the decision boundaries ('buffer' between hyperplane and nearby 'support vectors') | Use gradients of a loss function (e.g., *hinge loss*) to update weights $w_i$'s, thus iteratively maximizing margin |
| **Unsupervised** (only unlabelled data) | $k$-means clustering (other example families of methods include hierarchical clustering and dimensionality reduction approaches) | For a collection of unlabeled data, creates $k$ 'natural' grouping (or sets of associations) between the entities in the set | Gives set of maximal distances between $k$ centroids in unlabeled dataset (maximizes sum of squared *between*- cluster distances, which is equivalent to minimizing sum of squared distances [to centroid] *within* each cluster); equivalently, partitions the dataset into Voronoi cells | Iterate between two stages: (i) assign point $x_i$ to nearest cluster (lowest distance to mean), (ii) re-compute means, given all pts assigned to each cluster, and then (iii) iterate to convergence (no further Δs to point assignments); a greedy algorithm that partitions into $k$ groups |

| Type of ML (category, or "learning style") | Sample algorithms | Task | Performance | Experience |
|---|---|---|---|---|
| **Reinforcement learning** (either *model-based* or *model-free*, as in *Q*-learning) | *Q*-Learning (other RL methods include temporal difference learning, deep-*Q* networks [DQN], actor/critic framework, associative RL) | Determines a *policy* (transition probabilities for state/action pairs), that is optimal in sense of maximizing the expected cumulative (final) *reward*, given current state | *Q* is the *action-value* function that is iteratively optimized (policy updates via Bellman equation); in so doing, many parameters can be adjusted (*learning rate*, *discount factor*, initialization values [$Q_0$], etc.); in RL, the performance (and learning) occurs 'online'/on-the-fly | Classic way is to iteratively improve *Q* by up-dating via the Bellman equation for optimization (via *dynamic programming*); this recursive principle is that the optimal policy at state *i*+1 must subsume optima up to state *i*. |
| **Other, related approaches** | Genetic algorithms (GAs) | Find approximate solutions to an optimization problem, generally occupying a rather high-dimensional parameter space; the *solutions*, or individuals comprising the population, are encoded as *chromosomes* (e.g., as bit strings with blocks of 'genes'). | A given trial solution(/individual) is evaluated against an objective function (*fitness function*); notably, in GAs these functions can have mathematical properties that challenge traditional numerical optimization approaches (e.g., discontinuous, non-differentiable, highly nonlinear). | A subset of the *fittest individuals* (chromosomes yielding optimal values against the fitness function) are selected, along with a randomly chosen subset; these parents reproduce via operations like crossover (splicing chromosomes), mutations, etc. (biological evolution). Thus, the *population* of individuals evolves towards higher fitness, and solutions can be identified (as individual chromosomes). |
| | Swarm-based approaches, e.g. particle-swarm optimization, ant-colony optimization, dragonfly optimization | Estimate parameters, in a high-dimensional parameter space, that optimizes a global objective function (e.g., shortest cumulative path between two points, in ant-colony optimization). Sets of parameters are encoded as attributes of individual agents (particles, ants, etc.). | The population of individual entities (ants, agents, etc.) is evaluated against an optimality criterion (fitness function) defined by the modeler. Once a 'stopping criterion' is met, a solution to the task can be considered as optimal. | In a given iteration, agents (particles, ants, etc.) are updated (towards other ants, the centroid of a swarm, etc.) based on a combination of terms, one of which is a "social learning" parameter that is, itself, updated; crucially, this social parameter enables global communication amongst entities (e.g., as ant pheromones), and thus the population *collectively* equilibrates towards higher-fitness regions of the solution space. |

# Figures

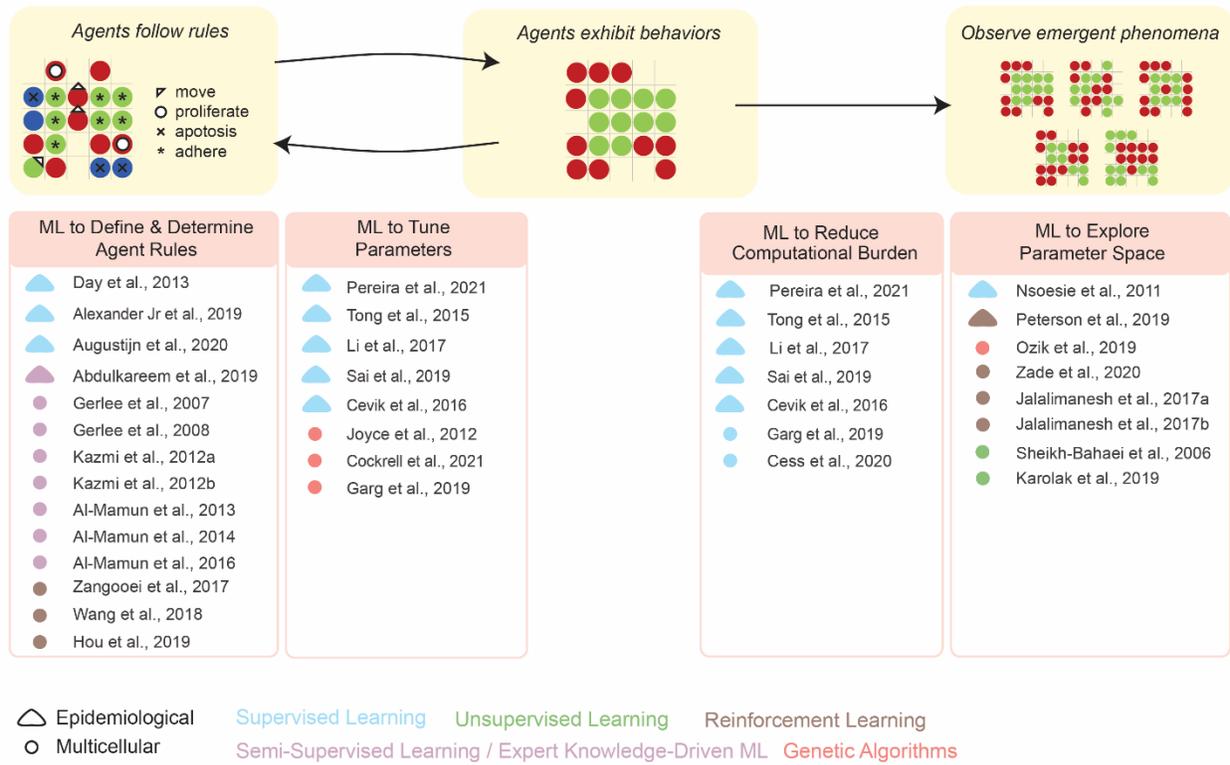

FIGURE 1. This overview schematizes how ML can aid the various stages in the development and application of an ABM—define/determine agent rules, tune parameters, explore parameter space, etc. Representative examples are given for various types of synergistic ML–ABM couplings (see literature citations).

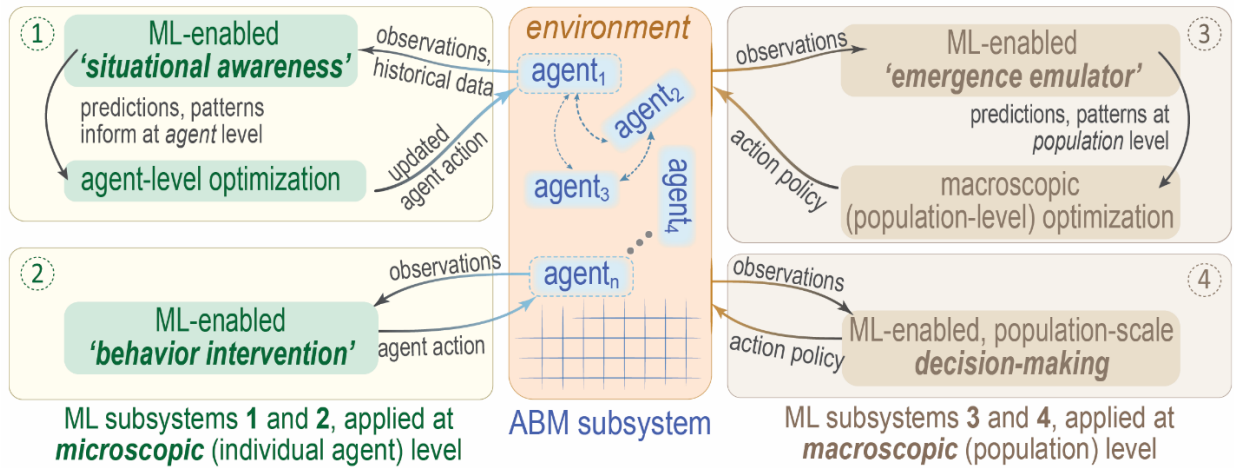

**FIGURE 2.** Schematic overview of potential ML/ABM integration schemes. This diagram suggests four modes (numbered circles) by which one might integrate ML and ABM, depending on whether the ML subsystem acts to optimize at the microscopic scale of individual agents (① and ②, left side) or else the macroscopic level of an entire population of agents (③ and ④, right side). Agents 1, 2, …, *n* are drawn in the ABM subsystem (middle panel), with coupling between agents denoted by dashed arrows; the square lattice is purely to emphasize the discrete nature of the ABM approach. As an example of how to interpret this diagram, note that mode ②, 'behavior intervention', entails application of 'online' ML methods to modify agent behavior/action policies, which is essentially reinforcement learning. Note that this illustration is adapted from one that appears in (Zhang et al., 2021), where further details may be found.

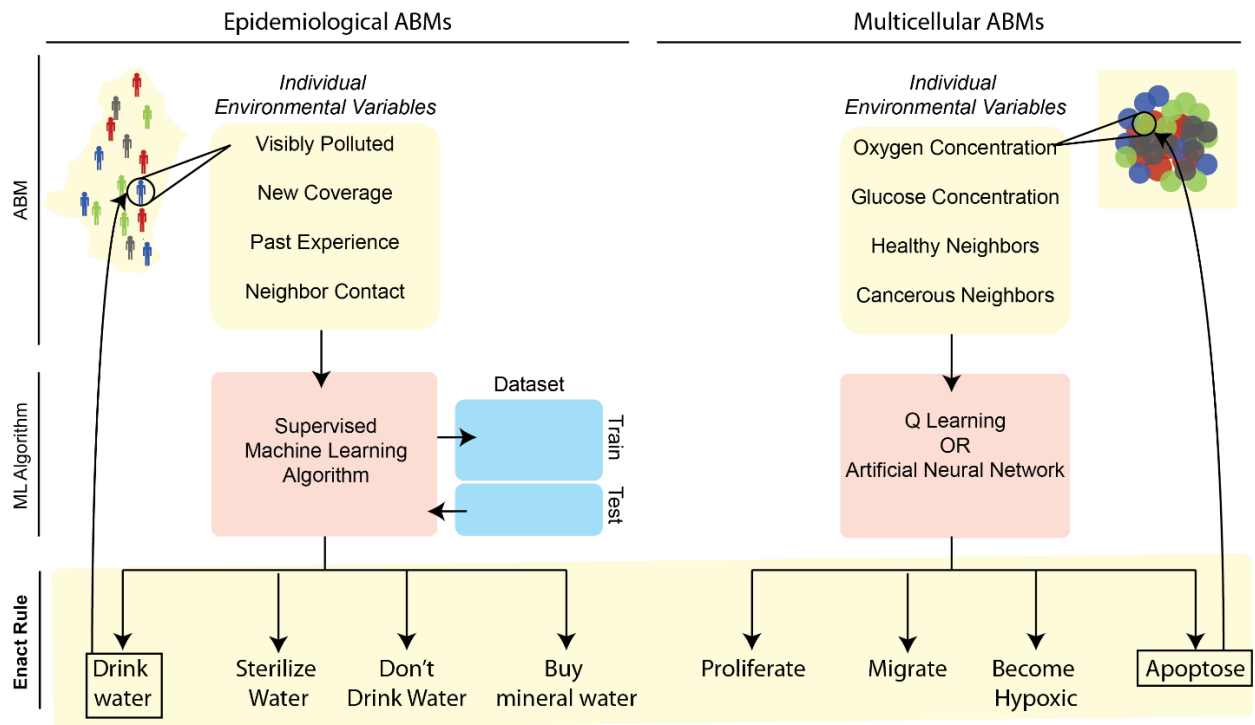

**FIGURE 3.** Application of ML to define rulesets in epidemiological (left) and multicellular (right) ABMs. In these two illustrative examples, ML-related stages are in red or blue while ABM-related steps are highlighted in yellow. In both contexts, individual agents survey environmental variables at a given time-step of the simulation. These environmental variables form the input for an ML algorithm that outputs a decision for the agent to enact. The left example refers to an ABM developed to simulate cholera spread in Kumasi, Ghana, wherein supervised learning algorithms trained on survey data were used to select the most probable behavior based on environmental variables (Abdulkareem et al., 2019; Augustijn et al., 2020). Several other epidemiological ABMs have leveraged large datasets to train supervised learning algorithms to determine agent behavior (Day et al., 2013; Abdulkareem et al., 2019; Alexander Jr et al., 2019; Augustijn et al., 2020). The right-hand example references multicellular ABMs that simulate individual cell behaviors in a tissue, with cellular 'decisions' being made based on either *Q*-learning (Zangooei and Habibi, 2017) or ANN approaches (Gerlee and Anderson, 2007; 2008; Kazmi et al., 2012a; Kazmi et al., 2012b; Al-Mamun et al., 2013; Al-Mamun et al., 2014; Al-Mamun et al., 2016; Abdulkareem et al., 2019).

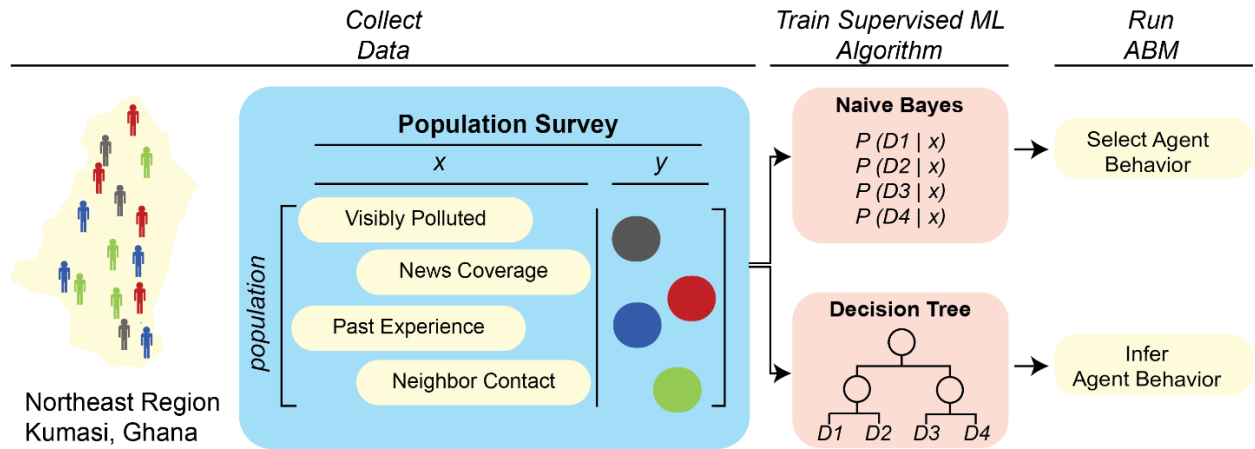

FIGURE 4. **Contrasting two applications of supervised learning algorithms to define agent rulesets in epidemiological ABMs**. Two studies applied ML to define agent rulesets in an ABM developed to simulate cholera spread in Kumasi, Ghana. The first study (upper pathway) applied a naive Bayes model to predict water usage of individual agents based on environmental variables (Abdulkareem et al., 2019), while the second one (lower path) trained a decision tree to derive agent behavior based on the same environmental variables (Augustijn et al., 2020). The two ML–ABM integrations predicted different numbers of total infected individuals in the population, illustrating that the ML algorithm used to adaptively refine an agent's behavior can impact the overall system-wide trends predicted by an ABM.

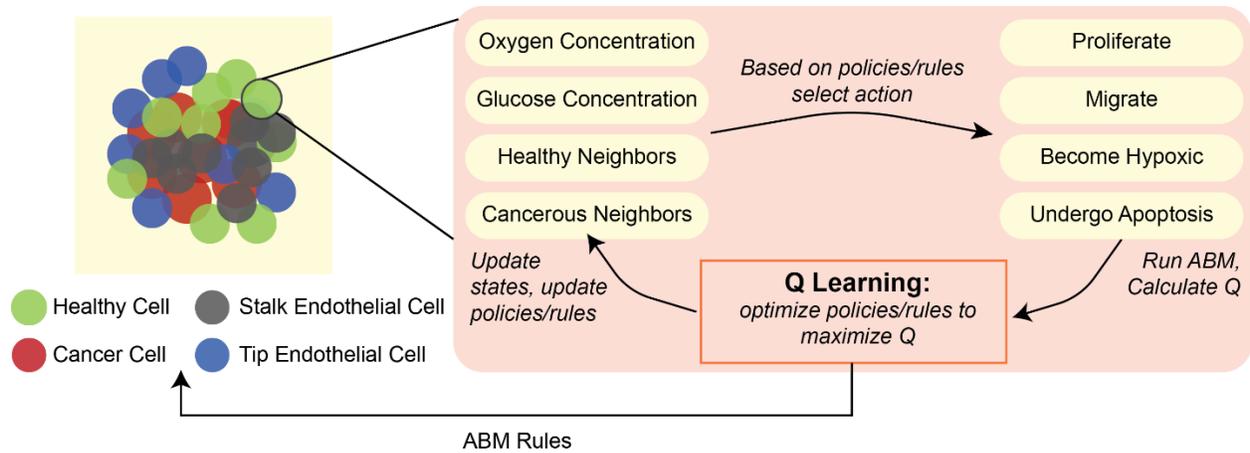

FIGURE 5. **Application of *Q*-learning to update cellular states in an ABM of tumorigenesis**. An ABM of 3D tumorigenesis (Zangooei and Habibi, 2017) applied *Q*-learning to find the optimal cell actions (proliferate, migrate, become hypoxic, undergo apoptosis) based on an individual cell's surrounding environmental variables, including oxygen concentration, glucose concentration, number of healthy cell neighbors, and number of cancerous cell neighbors.

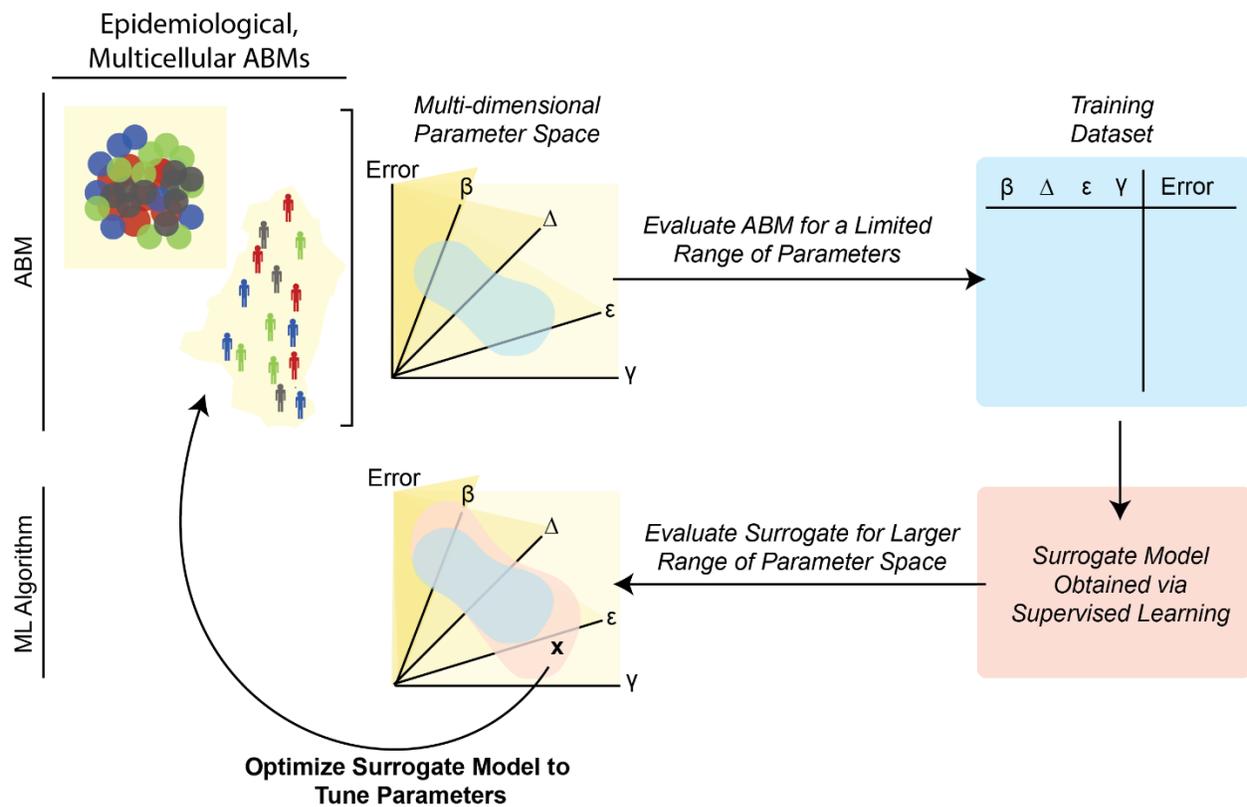

FIGURE 6. **Training surrogate ML models can reduce the computational burden of ABM calibration.** Because it requires repeatedly evaluating potentially complex numerical expressions, for multiple agents over numerous timesteps, ABM parameterization can be computationally expensive. Once trained (i.e., as applied in the inference stage), ML models are generally less computationally costly because they entail evaluating a single function to generate a prediction. Several studies (Tong et al., 2015; Cevik et al., 2016; Li et al., 2017; Sai et al., 2019; Pereira et al., 2021) have leveraged this advantage of ML by first evaluating an ABM for a limited range of parameters, and then creating a 'surrogate' dataset that relates the ABM parameters to final error. Next, a surrogate supervised learning algorithm can be trained on this data, and the surrogate model can then be used to explore broader regions of the original ABM's parameter space.

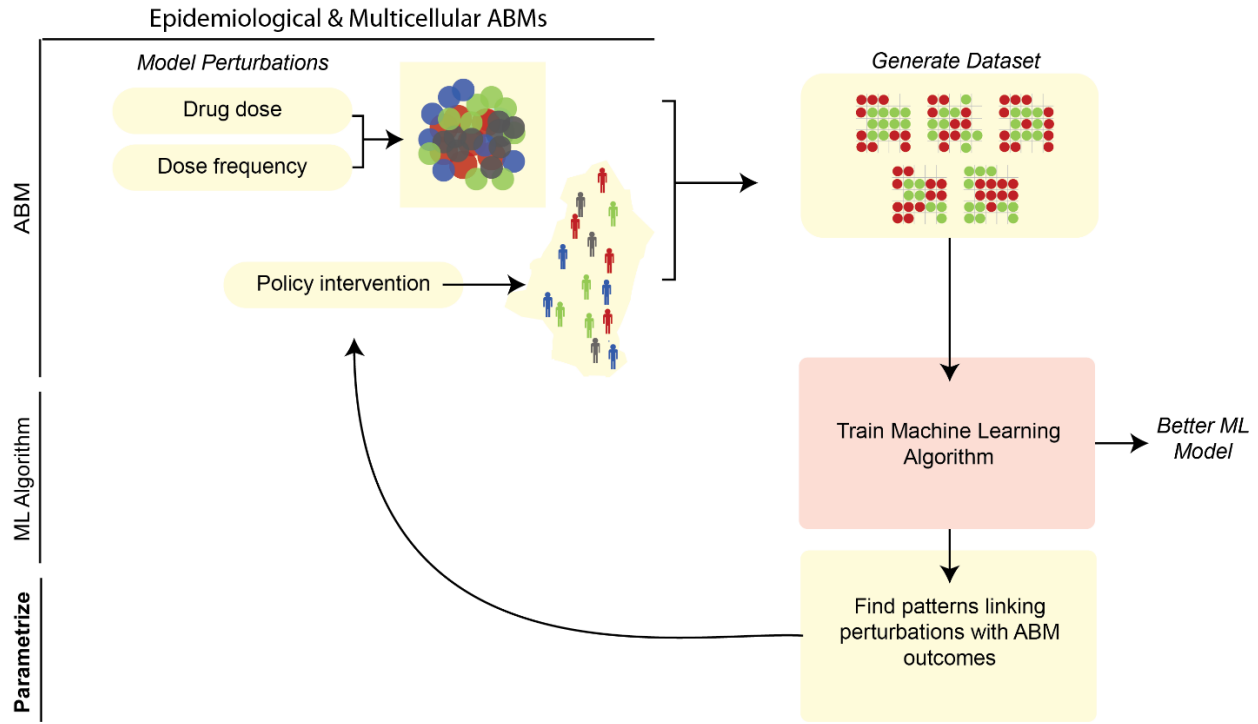

**FIGURE 7. ML can be applied to parameter space exploration of ABMs.** ABMs can be used to generate vast volumes of data and explore how system perturbations affect population-level outcomes in the model. After generating simulation data from an ABM, ML can be used to characterize patterns in the ABM (Karolak et al., 2019). Simultaneously, the datasets generated by ABMs can be used to train more robust ML algorithms (Sheikh-Bahaei and Hunt, 2006; Nsoesie et al., 2011).

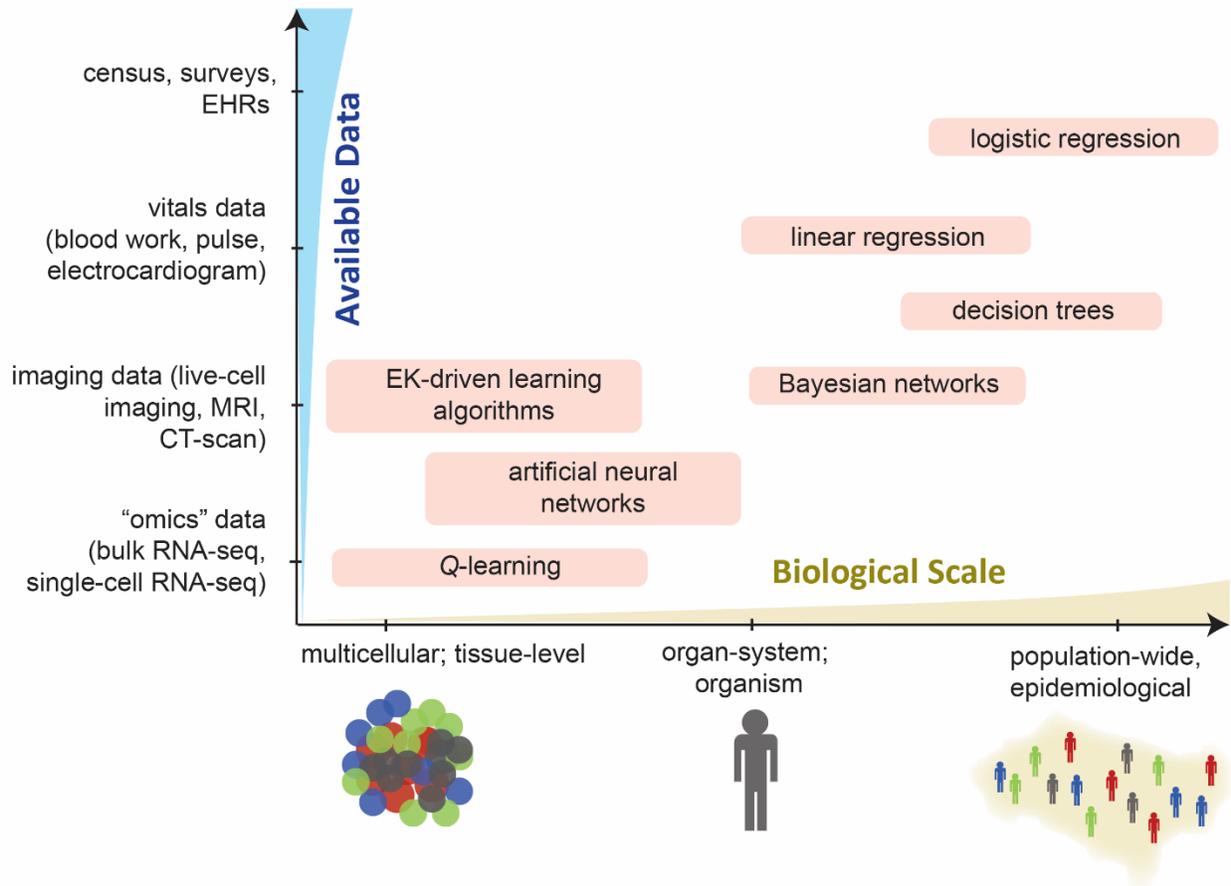

FIGURE 8. Integrated ABM/ML approaches depend on system scale and data characteristics. This highly schematic representation suggests how the suitable ML method to integrate with a particular ABM approach might vary with the scale of the biological system (horizontal axis) as well as the properties of the available datasets (vertical axis), such as volume, variety, and so on.